\documentclass[pra,twocolumn,superscriptaddress,showpacs,groupedaddress]{revtex4-1}
\usepackage{hyperref}
\usepackage{physics}
\usepackage{dsfont}
\usepackage{amsfonts}
\usepackage{bm}
\usepackage{graphicx}
\usepackage{xcolor}
\usepackage{mathtools}
\usepackage{verbatim}
\usepackage{tikz}
\usepackage[shortlabels]{enumitem}

\usepackage{dsfont}
\usepackage{bbold}

\begin{document}
	
	\title{Fate of entanglement in quadratic Markovian dissipative systems
	}
	\date{\today}
	\author{Fabio Caceffo}
	\affiliation{Dipartimento di Fisica dell'Universit\`a di Pisa and INFN, Sezione di Pisa, I-56127 Pisa, Italy}
	\author{Vincenzo Alba}
	\affiliation{Dipartimento di Fisica dell'Universit\`a di Pisa and INFN, Sezione di Pisa, I-56127 Pisa, Italy}
	
	\begin{abstract}
		We develop a hydrodynamic description for the driven-dissipative dynamics of the entanglement 
		negativity, which quantifies the genuine entanglement in mixed-state systems. 
		We focus  on quantum quenches in fermionic and bosonic 
		systems subject to linear dissipation, as described by quadratic Lindblad master equations. 
		In the spirit of hydrodynamics, we divide the system into mesoscopic cells. 
		At early times, correlations are generated in each cell 
		by the unitary component of the evolution. 
		Correlations are then transported across different cells 
		via ballistic quasiparticle propagation, 
		while simultaneously evolving under the action of the environment. 
		We show that  in the hydrodynamic limit  the negativity can be reconstructed from 
		the correlations between the independently propagating quasiparticles. 
		We benchmark our approach considering quenches from both homogeneous and inhomogeneous initial states 
		in the Kitaev chain, the tight-binding chain, 
		and the harmonic chain in the presence of gain/loss dissipation. 
	\end{abstract}
	
	\maketitle
	
\section{Introduction}
	\label{sec:intro} 

	In out-of-equilibrium closed quantum many-body systems, entanglement is intricately 
	intertwined with thermodynamics. This connection is particularly evident 
	in integrable systems, where the complete dynamics of entanglement 
	over time can be fully reconstructed~\cite{fagotti2008evolution,alba2017entanglement,alba2018entanglement} 
	from the Generalized Gibbs Ensemble ($GGE$), 
	which characterizes the \emph{local} properties of the steady state. 
	This is surprising because entanglement reflects ``nonlocal'' 
	correlations within the system, which cannot be directly encoded in the $GGE$. 
	Understanding the fate of entanglement and its relationship with thermodynamics in 
	open quantum systems is of paramount importance because every 
	realistic system interacts with an external environment. 
	Unfortunately, the dynamics of 
	open systems leads towards a globally mixed state, 
	rendering popular entanglement measures for pure states, such as the 
	von Neumann entropy, ineffective, due to contamination by 
	environment fluctuations. The logarithmic negativity~\cite{vidal2002negativity,plenio2005logarithmic,calabrese-2012,shapourian2019entanglement}, 
	an appropriate measure of entanglement even for mixed states, 
	proves challenging to compute, except in very specific scenarios~\cite{alba2022logarithmic,murciano2023symmetryresolved,caceffo2023entanglement}. 
	In contrast, in closed integrable systems the 
negativity is understood~\cite{alba2019quantum,murciano2021quench,kudler2020correlation} within the 
	quasiparticle picture for 
	entanglement~\cite{calabrese2005evolution,fagotti2008evolution,alba2017entanglement,calabrese2018entanglement,alba2021generalized,klobas2021exact,klobas2021exact,bertini2022growth}.

	Here we develop a hydrodynamic framework for the logarithmic negativity  
	between two intervals of equal length $\ell$ and placed 
	at distance $d$ (see Fig.~\ref{fig:numdata} (a))
	in generic \emph{quadratic} fermionic and bosonic systems \emph{\`a la} 
	Lindblad~\cite{petruccione2002the}. 
	Exact results in noninteracting systems proved to be crucial for 
	understanding the absence of thermalization~\cite{calabrese2011quantum} in integrable systems, 
	provided a valuable platform to test Generalized Hydrodynamics~\cite{alba2021generalized} 
	($GHD$), and  pioneered 
	the application of the quasiparticle picture~\cite{fagotti2008evolution}. 
	Free systems also facilitate the exploration of dynamics beyond the quantum 
	quench paradigm~\cite{prosen2011nonequilibrium}. 
	Quadratic Lindblad dynamics 
	are exactly solvable~\cite{prosen2008third}, 
	enabling a comprehensive understanding of 
	the dynamics of von Neumann and R\'enyi 
	entropies~\cite{alba2021spreading,Carollo2022entdiss,Alba2022entdiss,alba2022noninteracting,alba2021unbounded,caceffo2023entanglement,starchl2022relaxation}.

	Our framework is predicated on the idea that, at a hydrodynamic level, 
	the system can be partitioned into ``mesoscopic'' cells, which 
	are thermodynamically large relative to any microscopic 
	length scales, yet small in comparison to the overall 
	system size. At early stages of the dynamics, entanglement between 
	quasiparticles generated within each cell is established by the local unitary 
	component of the evolution. Subsequently, correlation and entanglement 
	are ``transported'' through the ballistic quasiparticle propagation. 
	Similar concepts have been previously explored for 
	unitary dynamics~\cite{bertini2018entanglementevolution,bastianello2018spreading}. 
	Crucially, the quasiparticles are 
	constructed from the \emph{full} Liouvillian generating the dynamics. 
	Indeed, as correlations spread 
	across different cells, they become ``dressed'' by the environment, 
	which renders them time-dependent. Generally, even with weak dissipation, 
	entanglement is exponentially dampened (see Fig.~\ref{fig:numdata} (b)). Bosonic systems experience the 
	so-called sudden death, wherein  
	the negativity is exactly zero at long times. Unlike in unitary dynamics, 
	where the negativity is linked to thermodynamic entropy~\cite{alba2019quantum}, 
	here it depends on ``off-diagonal'' 
	correlations between quasiparticles, rather than solely on their densities. 
	This contrasts with the behavior of von Neumann and R\'enyi 
	entropies in quadratic Liouvillians~\cite{Carollo2022entdiss}. 
	Thus, the dissipative dynamics profoundly reveals the breakdown 
	of the relationship between entanglement and thermodynamics. 
	The manuscript is organized as follows. In Section~\ref{sec:hydro-frame} we outline the 
	hydrodynamic framework for the negativity in generic quadratic fermionic and bosonic systems. 
	In Section~\ref{sec:fermions} we apply the method to derive the dynamics of the 
	fermionic negativity in paradigmatic fermion systems, such as the Kitaev chain and the tight 
	binding chain in the presence of gain and loss dissipation. In Section~\ref{sec:bosons}, {as an example for the bosonic case,} 
	we discuss the negativity dynamics in the so-called harmonic chain. In Section~\ref{sec:bipartite} 
	we discuss dynamics in the tight-binding chain starting from bipartite initial states in 
	the presence of homogeneous gain and loss dissipation. We conclude and discuss future directions 
	in Section~\ref{sec:concl}.
	{In Appendix~\ref{app:symm} we specify some mathematical details on the relation between physical symmetries and Fourier transforms of quadratic operators.}
	In Appendix~\ref{app:eigen} we discuss the effect of symmetries {of the Hamiltonian and of the dissipation} on the definition of quasiparticles.
	In Appendix~\ref{app:chbasis} we present a hydrodynamic derivation of our main results. In Appendix~\ref{app:detail} we 
	provide a detailed derivation of our {analytic} results for the negativity in the bosonic and fermionic 
	models that we consider. 

	\section{Hydrodynamic framework for the negativity}
	\label{sec:hydro-frame}

	Let us consider a generic free-fermion or free-boson Hamiltonian $H$ that is diagonalized in terms of $n$ species of 
	quasiparticles $\eta_j(k)$, with $j\in[1,n]$, and $\eta_j$ satisfying standard 
	commutation or anticommutation relations. The quasimomentum $k$ 
	lives in a reduced Brillouin zone $\mathcal{B}$. 
	In the spirit of hydrodynamics, let us divide the system 
	in mesoscopic cells.  A mesoscopic cell is thermodynamically large compared to the 
	microscopic length scales of the system, i.e., the lattice spacing for models defined on 
	a lattice. At the same time, each cell is small if compared to the full system. 
	After a quantum quench, 
	each cell can be viewed as source of the same excitations $\eta_j(k)$ 
	of the full chain. 
	\begin{figure*}
		\includegraphics[width=\linewidth]{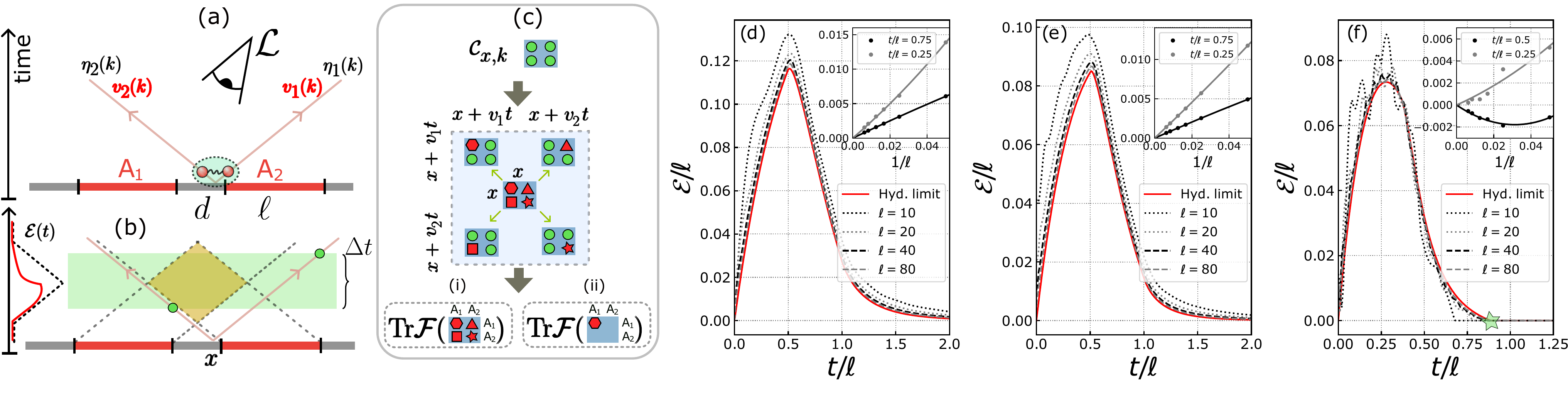}
		\caption{Dynamics of the negativity ${\cal E}(t)$ 
			between two equal intervals $A_1$ and $A_2$ of length $\ell$ 
		placed at distance $d$ under the Liouvillian 
			$\mathcal{L}$ in dissipative noninteracting systems. 
			(a) 
			Two entangled quasiparticles $\eta_1,\eta_2$ originated in the same hydrodynamic cell  
			propagate with velocities $v_1(k)$ and $v_2(k)$. 
			(b) The pair entangles $A_1$ and $A_2$ only in the time interval $\Delta t$, 
			in which it  is shared between $A_1$ and $A_2$. 
			The number of shared pairs is the horizontal width of the shaded diamond. 
			In the absence of dissipation, ${\mathcal E}(t)$ 
			exhibits a linear ``rise and fall'' dynamics, whereas  with dissipation 
			the ``rise and fall'' is exponentially damped.   
			(c) Entanglement in the hydrodynamic cell at $x$  is encoded in  the  $4\times 4$ correlator  
			${\mathcal{C}}_{x,k}$ (cf.~\eqref{eq:corr} and~\eqref{eq:trace-hydro-diss}) 
			between the quasiparticles. At time $t$, the 
			``transported'' correlator  $\widetilde{\mathcal{C}}_{x,k,t}(t)$  
			describes the negativity between quasiparticles in the cells at 
			$x+v_1t$ and $x+v_2t$. If $x+v_1t\in A_1$ and $x+v_2t\in A_2$ 
			the contribution to the negativity between $A_1$  and $A_2$ is reported in (i), 
			whereas (ii) shows the configuration with a quasiparticle in $A_1$, which gives  
			${\cal E}=0$. (d-f). Numerical benchmarks for gain/loss dissipation with rates $\gamma_\pm$. 
			${\mathcal E}(t)/\ell$  versus $t/\ell$ in the weakly dissipative hydrodynamic limit. 
			(d) Quench from the dimer state in the tight-binding chain with $\gamma_+\ell=0.2$, $\gamma_-\ell=0.3$. (e) Magnetic field quench with 
		        $h_0=0.3$, $h_1=3.1$, $\delta=0.73$ in the Kitaev chain. The dissipation rates are $\gamma_+\ell=0.2$, $\gamma_-\ell=0.5$. (f) Mass quench with 
			$m_0=1.3$, $m_1=0.5$, $w=1.73$, $\mathcal{K}=2.1$, $\gamma_+\ell=0.25$, $\gamma_-\ell=0.9$ in the harmonic chain. 
			The full lines are analytic results. The insets show  
			finite-size corrections at fixed $t/\ell$. Curves are fits to $a/\ell+b/\ell^2$. 
			In (f), the star marks the negativity sudden death. 
		}
		\label{fig:numdata}
	\end{figure*}
	For now, to illustrate the hydrodynamic framework, let us focus on  
	the unitary dynamics under $H$, starting from {a Gaussian} initial state. 
	{Because of the Gaussianity,}
	the correlations initially generated  in the cell at position $x$ are characterized 
	by the $2n\times 2n$ matrix $\mathcal{C}_{x,k}$ as 
	\begin{equation}
	\label{eq:corr}
	\mathcal{C}_{x,k}=\left\langle 
	\begin{pmatrix}
	\eta^\dagger_r(k)\eta_s(k) & \eta^\dagger_r(k)\eta_s^\dagger(k)\\
	\eta_r(k)\eta_s(k) & \eta_r(k)\eta^\dagger_s(k)
	\end{pmatrix}
	\right\rangle,\quad r,s\in[1,n], 
	\end{equation}
	with $\langle\cdot\rangle$ the initial-state 
	expectation value.  ${\mathcal C}_{x,k}$ is 
	block diagonal in momentum space for translational-invariant $H$ and homogeneous initial states, 
	implying that one could drop the dependence on $x$ in~\eqref{eq:corr} because all the cells are equal.
	Crucially, $\mathcal{C}_{x,k}$ is not diagonal as $2n\times2n$ matrix, 
	reflecting that the $n$ quasiparticles for each $k$ form an entangled \emph{multiplet}. 
	We consider dynamics that preserve the Gaussianity of the initial states, 
	implying that one can univocally associate  to 
	${\mathcal C}_{x,k}$ a Gaussian density matrix  $\rho_{x,k}$. 
	The full-system correlation matrix is $\mathcal{C}=\bigoplus_{x,k}
	\mathcal{C}_{x,k}$, and the density matrix $\rho=\bigotimes_{x,k}\rho_{x,k}$, 
	since different cells are decoupled. 
	Notice that the change of basis from the lattice fermions or bosons to the 
	$\eta_j(k)$ in each cell can be subtle and is discussed in 
	Appendix~\ref{app:eigen}. 
	Quasiparticles of each multiplet spread ballistically with group velocities 
	$v_j(k)$, which are obtained from the single-particle dispersion $\varepsilon(k)$ of the model. 
	Correlations spread accordingly, via the mechanism highlighted in 
	Fig.~\ref{fig:numdata} (c). 
	At time $t$, the $j$-th quasiparticle of the multiplet created in the cell at $x$ 
	is in the cell at $x+v_j(k)t$. Suggestively, we can associate to each multiplet a 
	non-local correlation matrix ${\mathcal C}_{x,k,t}$ (notice the subscript $t$) obtained by ``transporting'' 
	$\mathcal{C}_{x,k}$ (cf.~\eqref{eq:corr}). $\mathcal{C}_{x,k,t}$ describes correlations 
	between cells at $x+v_jt$, $j=1,...,n$, although its entries  are  simply those of ${\mathcal C}_{x,k}$. 
	We illustrate  the case with $n=2$ 
	in Fig.~\ref{fig:numdata} (c), where  the triangle and the square are 
	the $2\times 2$ correlation blocks (cf.~\eqref{eq:corr}) 
	between $\eta_1$ in the cell at $x+v_1t$ and $\eta_2$ in the cell at $x+v_2t$. 
	The hexagon and the star are the correlations $\langle\eta_{j}^{\scriptscriptstyle(\dagger)}\eta_{j}^{\scriptscriptstyle(\dagger)}\rangle$  
	inside the same cell at $x+v_{j}t$.  Again, ${\mathcal C}_{x,k,t}$ identifies 
	the Gaussian state $\rho_{x,k,t}$, the full-chain correlation matrix being  
	${\mathcal C}_{t}=\bigoplus_{x,k}{\mathcal C}_{x,k,t}$, 
	whereby the full-system state is $\rho_t=\bigotimes_{x,k}\rho_{x,k,t}$. 
	
	Let us now consider the logarithmic negativity ${\cal E}$ between two intervals $A_1$ and $A_2$ 
	(see Fig.~\ref{fig:numdata} (b)). 
	${\cal E}$  is obtained as $f(\rho_A)$, with $\rho_A$ the reduced 
	density matrix of $A=A_1\cup A_2$ and 
	$f(x)$ a suitable function~\cite{shapourian2019entanglement,Audenaert2002negativity}. 
	Gaussianity implies that 
	$f(\rho_A(t))=\mathrm{Tr}_A \mathcal{F}(\mathcal{C}_t^{\scriptscriptstyle(A)})$, 
	with $\mathcal{C}_t^{\scriptscriptstyle(A)}$ being $\mathcal{C}_t$ restricted to $A$, 
	and $\mathcal{F}(x)$ a function~\cite{shapourian2019entanglement,Audenaert2002negativity}, 
	which is related to $f(z)$. Our hydrodynamic-limit assumptions imply 
	\begin{equation}
	\label{eq:trace-hydro}
	f(\rho_A(t))
	\stackrel{\mathrm{hydr}}{=}
	\sum_{x,k}\mathrm{Tr} \mathcal{F}(\mathcal{C}^{\scriptscriptstyle(\mathcal{A})}_{x,k,t}). 
	\end{equation} 
	In~\eqref{eq:trace-hydro}, $\mathcal{C}_{x,k,t}^{\scriptscriptstyle(\mathcal{A})}$ is 
	$\mathcal{C}_{x,k,t}$, where we select the rows and columns  
	corresponding to the quasiparticles, with labels $\mathcal{A}\subseteq \{1,2,\dots,n\}$,  
	that at time $t$ are in  $A$. 
	The sum over $x,k$ depends on the way quasiparticles of the same multiplet distribute 
	between $A_1$ and $A_2$. Eq.~\eqref{eq:trace-hydro} holds  in the 
	hydrodynamic limit $\ell,t,d\to\infty$ with $\ell/t$, $d/t$ fixed.
	For $n=2$ (entangled pairs, see Fig.~\ref{fig:numdata} (a,b)), one has 
	quite generically $v_1(k)>0, v_2(k)=-v_1(k)$, with $k\in[0,\pi]$. 
	Moreover, for a generic 
	entanglement measure, $\mathcal{F}(\mathcal{C}_{x,k,t}^{\scriptscriptstyle(\mathcal A)})$ (cf.~\eqref{eq:trace-hydro}) 
	is nonzero only if the pair is shared between $A_1$ and $A_2$ (Fig.~\ref{fig:numdata} (c)).
	Thus, Eq.~\eqref{eq:trace-hydro} gives 
	\begin{equation}
	\label{eq:neg_leading}
	\mathcal{E}(t)=\ell \int_{0}^{\pi}\frac{dk}{2\pi} e_2(k,t) \Theta(k,t),
	\end{equation}
	where $e_2(k,t)=\mathrm{Tr}(\mathcal{F}(\mathcal{C}^{\scriptscriptstyle(\mathcal{A})}_{x,k,t}))$,
	and
	\begin{multline}
	\label{eq:time_theta}
	\Theta(k,t)=\max(2v_1t/\ell,2+d/\ell)+\\
	\max(2v_1t/\ell,d/\ell)-2\max(v_t/\ell,1+d/\ell)
	\end{multline}
	results from the sum over $x$ in~\eqref{eq:trace-hydro}, and it gives  
	the linear ``rise and fall'' dynamics (see Fig.~\ref{fig:numdata} (b)). 
	For unitary dynamics  
	$\rho_{x,k,t}$ corresponds to a pure state for the pair, 
	implying that ${\cal E}=1/2 I_{A_1:A_2}^{\scriptscriptstyle{(1/2)}}$, with 
	$I_{A_1:A_2}^{\scriptscriptstyle{(1/2)}}$ the R\'enyi-$1/2$ mutual 
	information~\cite{alba2019quantum}. In the presence of dissipation $e_2(k,t)$ 
	will depend on time, and the rise and fall curve will be exponentially dampened (see Fig.~\ref{fig:numdata} (b)). 
	
	To deal with dissipation, we employ the Lindblad master equation~\cite{petruccione2002the}. 
	The evolution of a generic observable $O$ is given as 
	$d\langle O\rangle/dt=i\langle\hat{\mathcal{L}}^\dagger[O]\rangle$, with 
	\begin{equation}
	\label{eq:lindblad}
	\hat{\mathcal{L}}^\dagger[O]=[H,O] -i \sum_{j,j'=1}^{2L} K_{jj'} 
	\big(r_j O r_{j'} -\frac{1}{2}\left\{r_j r_{j'},O \right\}\big),
	\end{equation}
	where $\langle O\rangle:=\mathrm{Tr}(O\rho_t)$ and $L$ is the length of the chain (in our framework, we take the thermodynamic limit $L\to\infty$). The Kossakowski matrix $K$ encodes the dissipation~\cite{lindblad1976on}, 
	and it can be decomposed as $K=A+iB$, with $A (B)$ real symmetric (antisymmetric) matrices. 
	For fermionic systems, $r=(w_1^1,w_1^2,\dots,w_{L}^1,w_L^2)$ are Majorana fermions, 
	whereas for bosons $r=(x_1,p_1,\dots,x_L,p_L)$, with $x_j,p_j$ satisfying  
	$[x_j,p_{j'}]=i \delta_{jj'}$. 
	
	To proceed, we apply Eq.~\eqref{eq:lindblad} to each hydrodynamic cell. 
	First, we identify the quasiparticles $\eta_j$ transporting the 
	correlations as linear combinations of $r_n$~\cite{alba2021spreading,Carollo2022entdiss,starchl2022relaxation}. 
	For bosonic systems, the $\eta_j$ themselves are eigenvectors of $\hat{\mathcal{L}}^\dagger$. 
	For fermions, $\hat{\mathcal{L}^\dagger}[r_n]$  gives rise to nonlinear terms, 
	and to define the $\eta_j$ one has to consider the quadratic operators 
	$\chi(k)=\eta_j^\dagger\eta_j,\eta_i\eta_j$ requiring that they satisfy 
	the generalized eigenvalue equation (see Appendix~\ref{app:eigen} and Ref.~\cite{starchl2022relaxation})
	\begin{equation}
	\label{eq:L-evol}
	\hat{\mathcal{L}}^\dagger[\chi]=\lambda_\chi
	\chi(k)+\gamma_\chi,  
	\end{equation}
	where $\lambda_\chi(k)$, $\gamma_\chi(k)\in\mathds{C}$. 
	The group velocities of the quasiparticles are extracted 
	from $\chi=\eta_i\eta_j$ with $i\ne j$ (cf. Appendix~\ref{app:eigen} and Ref.~\cite{starchl2022relaxation}). 
	Although Eq.~\eqref{eq:L-evol} holds for finite dissipation, 
	we focus on the weakly-dissipative limit~\cite{alba2021spreading}, 
	in which $t,\ell\to\infty$, the dissipation rates 
	$\gamma$ vanish, and  $t/\ell$ and $\gamma\ell$ are fixed. 
	In this limit, the $\eta_j$ obey canonical 
	(anti)commutation relations (see Appendix~\ref{app:eigen}), and the quasiparticle velocities 
	$v_j(k)$ are the same as without dissipation (see Ref.~\cite{alba2021spreading,Carollo2022entdiss}). 
	Finally, we conjecture that Eq.~\eqref{eq:trace-hydro} generalizes as 
	\begin{equation}
	\label{eq:trace-hydro-diss}
	f(\rho_A(t))\stackrel{\mathrm{w.d.\,hydr}}{=}
	\sum_{x,k}\mathrm{Tr}\mathcal{F}(\widetilde{\mathcal{C}}^{\scriptscriptstyle(\mathcal{A})}_{x,k,t}(t)).
	\end{equation}
	Eq.~\eqref{eq:trace-hydro-diss} can be justified within the hydrodynamic framework, and 
	it relies on the independence of different hydrodynamic cells, and on the canonical commutation 
	relations between the quasiparticles (see Appendix~\ref{app:chbasis}). 
	To build $\widetilde{\mathcal{C}}_{x,k,t}(t)$, we first evolve~\eqref{eq:corr} 
	with~\eqref{eq:lindblad}, obtaining $\widetilde{\mathcal{C}}_{x,k}(t)$. Then, we transport 
	ballistically $\widetilde{\mathcal{C}}_{x,k}(t)$ as in the unitary case (see Fig.~\ref{fig:numdata}(c)), 
	to obtain $\widetilde{\mathcal{C}}_{x,k,t}(t)$. Importantly, in doing so we set to $1$ the 
	oscillating phases due to the unitary part of the dynamics, which in the hydrodynamic limit has already been accounted for by the quasiparticle spreading. Still, $\widetilde{\mathcal{C}}_{x,k,t}(t)$
	depends explicitly on time due to the dissipation, unlike in the unitary case. 
	We expect Eq.~\eqref{eq:trace-hydro-diss} to hold at finite dissipation rate as well, 
	at least for some types of dissipations (see Ref.~\cite{starchl2022relaxation}). For finite $\gamma$ 
	the eigenmodes of the Liouvillian $\eta_j(k)$ are related to those of the Hamiltonian by 
	a non-unitary matrix $V_k$ (see Appendix~\ref{app:eigen}). 
	The $\eta_j(k)$ satisfy the noncanonical anticommutation relations 
	$\{\eta_j(k),\eta_j^\dagger\}=V_k^\dagger V_k$. Now, 
	to generalize~\eqref{eq:trace-hydro-diss} one {would have to} take into account the effect of 
	$V_k$, as in Ref.~\cite{starchl2022relaxation}. Moreover, the velocities of the quasiparticles derived from the $\lambda_\chi$ (cf. Eq.~\eqref{eq:L-evol}) would not coincide with the unitary case.

\section{Applications to fermionic systems} 
	\label{sec:fermions}

\subsection{Field quench in the Kitaev chain with gain and loss dissipation}

	Let us consider the periodic Kitaev chain, defined by the  Hamiltonian
	\begin{equation}
		\label{eq:ham-kitaev}
		H_{\mathrm{K}}=-\frac{1}{2}\sum_{j=1}^{L} \big(c_j^\dagger c_{j+1}+\delta 
	c_j^\dagger c_{j+1}^\dagger + h. c. -2h c_j^\dagger c_j\big), 
	\end{equation}
	where $h,\delta$ are real parameters, and we take the thermodynamic limit $L\to \infty$.
	The strategy to diagonalize the Hamiltonian~\eqref{eq:ham-kitaev} is standard, and 
	we discuss it in Appendix~\ref{app:kitaev}. 
	The single-particle energies are 
	\begin{equation}
		\varepsilon(k)=[(h-\cos k)^2+\gamma^2\sin^2 k]^{1/2}. 
	\end{equation}
	In our quench protocol, at $t<0$ the chain is in the  ground state of $H_{\mathrm{K}}$  with $h=h_0$. 
	At {$t=0$}, $h$ is suddenly changed to $h=h_1$.  Pairs of entangled quasiparticles are formed by the 
	modes $\eta_1(k)$ and $\eta_2(k)$, with $k\in [0,\pi]$,  and 
	group velocities $v_1(k)=\partial_k \varepsilon(k)=-v_2(k)$ (see~\cite{alba2021generalized}). 
	We focus on the Lindblad dynamics in~\eqref{eq:lindblad}, where $r_{2m-1}=c_m+c_m^\dagger$, 
	$r_{2m}=i(c_m-c_m^\dagger)$. All the correlations are encoded in the matrix $\Gamma_{m,n}$ defined as 
	\begin{equation}
	\Gamma_{m,n}:=\langle [r_m, r_n]/2\rangle.
	\end{equation} 
	Its Fourier transform is 
	\begin{equation}
	\hat{g}_{k,a,b}=\langle r_{a,k} r_{b,-k}\rangle-\delta_{a,b},
	\end{equation} 
	with $a,b\in\{e,o\}$ and $r_{o,k}$, $r_{e,k}$ the Fourier transforms of $r_{2j+1}$ and $r_{2j}$, respectively. 
	Eq.~\eqref{eq:lindblad} for $\hat{g}_k$ gives~\cite{Carollo2022entdiss} 
	\begin{equation}
	\label{eq:corrt_k_ferm}
	\hat{g}_k=\hat{u}_k
	\hat{g}_k(0) \hat{u}_{-k}^T 
	+ 4i \int_{0}^{t} du \; \hat{u}_k(t-u) \hat{B}_k \hat{u}_{-k}^T(t-u),
	\end{equation}
	with $\hat{u}_k(t)=e^{-(4i\hat{h}_k+2\hat{A}_k)t}$, where $\hat{h}_k$,
	$\hat{A}_k$ and $\hat{B}_k$ are the Fourier transforms of $h_{m,n}$, which is defined as 
	$H_{\mathrm{K}}=\sum_{m,n}h_{m,n}r_mr_n$, and of $A,B$ (cf.~\eqref{eq:lindblad}). 
	We consider gain and loss dissipation with rates $\gamma_\pm$, which 
	corresponds to 
	\begin{align}
	& \hat{A}_k=((\gamma_++\gamma_-)/4) \mathds{1}_{2}\\
	&\hat{B}_k= i((\gamma_+-\gamma_-)/4) \sigma^y,
	\end{align} 
	with $\sigma^{x,y,z}$ the Pauli matrices. 
	For gain and loss the quasiparticles $\eta_j$ are the same as in the unitary case (cf. Appendix~\ref{app:kitaev}). 
	By solving~\eqref{eq:corrt_k_ferm} for $\hat{g}_k$~\cite{Alba2022entdiss}, 
	one obtains ${\widetilde{\mathcal C}}_{x,k}(t)$ as (see Appendix~\ref{app:kitaev})
	\begin{equation}
	\label{eq:corrt_xy}
	\widetilde{\mathcal{C}}_{x,k}(t)=\frac{1}{2}\begin{pmatrix}
	\widetilde{{\mathcal C}}_1&\widetilde{{\mathcal C}}_2^\dagger\\
	\widetilde{{\mathcal C}}_2& \widetilde{{\mathcal C}}_1
\end{pmatrix}. 
	\end{equation}
	Here 
	$\widetilde{\mathcal{C}}_1:=\mathds{1}_2-
	A_{k,t} \sigma^z, \quad \widetilde{\mathcal{C}}_2:=iB_{k,t}(e^{2i\varepsilon(k)t}\sigma^++h.c)$, where 
	\begin{align}
	\label{eq:AB}
	& A_{k,t}:=\lambda_t \cos (\Delta_{k,h_1}-\Delta_{k,h_0})
	\\&\nonumber\qquad+(1-\lambda_t)(1-2n_\infty)\cos \Delta_{k,h_1},\\
	\label{eq:AB-2}
	& B_{k,t}:=\lambda_t \sin (\Delta_{k,h_1}-\Delta_{k,h_0}),
	\end{align}
	with $\lambda_t:=e^{-(\gamma_+ + \gamma_-)t}$, $n_\infty:=\gamma_+/(\gamma_++\gamma_-)$, 
	and $\Delta_{k,h}$ the Bogoliubov angle (see Appendix~\ref{app:kitaev}). 
	We now set to $1$ the oscillatory terms $e^{\pm 2i\varepsilon(k)t}$ 
	in $\widetilde{\mathcal{C}}_2$, and we obtain $\widetilde{\mathcal{C}}_{x,k,t}(t)$ as in Fig.~\ref{fig:numdata} (c). 
	Finally,  we compute the negativity $e_n(x,k,t)$ between the quasiparticles that at time $t$ are 
	shared between $A_1$ and $A_2$, and which  form a multiplet originated in the cell at $x$, 
	as~\cite{Eisert_2018neg}  
	\begin{equation}
	\label{eq:neg-def}
	e_n(x,k,t)=\frac{1}{2}\left[S_{1/2}(\Gamma_\times(x,k,t))-
	S_2(\Gamma_{\mathcal{A}_1\cup\mathcal{A}_2})\right]
	\end{equation}
	where $S_\alpha$ are the R\'enyi-$\alpha$ entropies. 
	In~\eqref{eq:neg-def}, $\Gamma(x,k,t)=2 \widetilde{\mathcal{C}}_{x,k,t}(t)-
	\mathds{1}$, and  we defined $\Gamma_\times=[(\mathds{1}+\Gamma_{\mathcal{A}_1\cup\mathcal{A}_2}^2)/2]^{-1} 
	\mathrm{diag}(\Gamma_{\mathcal{A}_1},-\Gamma_{\mathcal{A}_2})$, 
	where $\mathcal{A}_1,\mathcal{A}_2$ contain the labels of the quasiparticles in $A_1$ and $A_2$, respectively. 
	In our quench we have $n=2$, and Eq.~\eqref{eq:neg_leading} holds with
	\begin{multline}
	\label{eq:xy_neg}
	e_2(k,t)=\ln\Big[\frac{1}{2} \left(1+A_{k,t}^2+B_{k,t}^2\right)+\\
	+\Big(\frac{1}{4}\left(1+A_{k,t}^2+B_{k,t}^2\right)^2-A_{k,t}^2\Big)^\frac{1}{2}\Big],
	\end{multline}
	where $A_{k,t}$, $B_{k,t}$ are defined in~\eqref{eq:AB} and~\eqref{eq:AB-2}. 

\subsection{Quench from the dimer state in the tight binding chain with gain and loss dissipation}

	Let us now consider the tight-binding chain, which 
	corresponds to $\delta=h=0$ in $H_{\mathrm{K}}$ (see Appendix~\ref{app:dimer}).  
	The Hamiltonian is diagonalized by going to momentum space as 
	\begin{equation}
		H=\int_0^\pi \frac{dk}{2\pi}\varepsilon(k)\left(c_k^\dagger c_k-c^\dagger_{k-\pi}c_{k-\pi}\right), 
	\end{equation}
	where $\varepsilon(k)=-\cos(k)$. 
	We focus on the quench from the dimer state 
	\begin{equation}
		|\mathrm{D}\rangle=2^{-L/4}\prod_{j=1}^{L/2}(c_{2j-1}^\dagger +c_{2j}^\dagger)
	|0\rangle.\end{equation} in the presence of gain and loss dissipation, which correspond 
	to Lindblad operators (in momentum space) $L_+\sim\sqrt{\gamma^+}c^\dagger_k$ and $L_-\sim 
	\sqrt{\gamma^-}c_k$.  {The quasiparticles species are now $\eta_1=c_k$ and 
	$\eta_2=c_{k-\pi}$ (see Appendix~\ref{app:dimer}), with the correlation matrix between them (cf.~\eqref{eq:corr}) reading}
\begin{multline}
\label{eq:corrt_neel_diss}
\widetilde{\mathcal{C}}_{x,k}(t):=\langle \eta^\dagger_i\eta_j\rangle= 
\mathcal{C}_{x,k}(t) e^{-(\gamma_++\gamma_-)t}+\\
\left(1-e^{-(\gamma_++\gamma_-)t}\right)\begin{pmatrix}
n_\infty&0&0&0\\
0&1-n_\infty&0&0\\
0&0&n_\infty&0\\
0&0&0&1-n_\infty
\end{pmatrix},
\end{multline}
where $n_\infty=\gamma^+/(\gamma^++\gamma^-)$, whereas $\mathcal{C}_{x,k}(t)$ is the 
unitarily evolved correlator 
\begin{widetext}
\begin{equation}
\label{eq:corrt_dimer_unitary}
\mathcal{C}_{x,k}(t)=\frac{1}{2}\begin{pmatrix}
1+\cos k&0&i\sin ke^{2i\varepsilon(k)t}&0\\
0&1-\cos k&0&i\sin k e^{-2i\varepsilon(k)t}\\
-i\sin k e^{-2i\varepsilon(k)t}&0&1-\cos k&0\\
0&-i \sin k e^{2i\varepsilon(k)t}&0&1+\cos k\\
\end{pmatrix}.
\end{equation}
\end{widetext}
	Now, Eq.~\eqref{eq:neg_leading} should hold 
	with  $e_2(k,t)=\ln(T_1+T_2)$, where $T_1,T_2$ are easily determined by applying~\eqref{eq:trace-hydro-diss} as 
	\begin{align}
	\label{eq:T-1}
	& T_1^{\mathrm{D}}:=\lambda_t+2g_\times(1-\lambda_t)^2\\
	\label{eq:T-2}
	& T_2^\mathrm{D}:=\big[(1-\lambda_t)^2[
	(1-\lambda_t)g_+ +\lambda_t]^2+\lambda_t^2\sin^2(k)\big]^{\frac{1}{2}}, 
	\end{align}
	with 
	$g_\times:=(\gamma^+\gamma^-)/(\gamma^++\gamma^-)^2$,   
	$g_+:=(\gamma^+)^2+(\gamma^-)^2/(\gamma^++\gamma^-)^2$
	and $\lambda_t$ as defined in~\eqref{eq:AB} and~\eqref{eq:AB-2}. 
	A similar result for the quench from the fermionic N\'eel state was 
	derived \emph{ab initio} in Ref~\cite{alba2022logarithmic}. The derivation 
	is cumbersome, while it is straightforward within our approach (see Appendix~\ref{app:tb-neel}). 
	
	\subsection{Numerical benchmarks}
	
	In Fig.~\ref{fig:numdata} (d-{e}) we discuss numerical 
	results for several Lindblad dynamics with gain and loss dissipation. In (d) 
	we consider dynamics in the tight-binding chain starting from the dimer state. 
	The dashed and dotted curves are exact numerical data, 
	and the continuous line is obtained by 
	using $e_2(k,t)=\ln(T_1+T_2)$  and~\eqref{eq:T-1}~\eqref{eq:T-2} in~\eqref{eq:neg_leading}. 
	We plot ${\cal E}/\ell$ versus $t/\ell$, considering the weakly-dissipative limit $\gamma_\pm\propto 
	1/\ell$. The agreement between 
	the numerical data and the quasiparticle picture is perfect in the limit $\ell\to\infty$. 
	The inset shows fits to $\sim a/\ell+b/\ell^2$ of the finite-size corrections. 
	In Fig.~\ref{fig:numdata} (e) we focus on the quench $h_0=0.3\to h_1=3.1$ 
	in the Kitaev chain. The agreement between the data and the analytic result  
	(cf.~\eqref{eq:neg_leading}~\eqref{eq:xy_neg}) is excellent.

\section{Applications to bosonic systems} 
	\label{sec:bosons}

	As a case-study of bosonic systems, we consider the mass quench 
	in the harmonic chain
	\begin{equation}
		H_{h}=\frac{1}{2}\sum_{j=1}^{L}\left
		(\frac{p_j^2}{m}+m w^2 x_j^2 + \mathcal{K}(x_{j+1}-x_j)^2\right),
	\end{equation} 
	where $x_j,p_j$ are canonically conjugated variables, and $\mathcal{K},w$ real parameters. 
	The initial state is the ground state 
	of $H_h$ with  $m=m_0$, that is 
	quenched to $m=m_1$. 
	Bosonic gain and loss dissipation corresponds to
	$\hat K_k=\hat A_k+i \hat B_k$ (cf.~\eqref{eq:lindblad}), with 
	$\hat{A}_k=(\gamma_++\gamma_-)\mathds{1}_{2}/2$ and  $\hat{B}_k= 
	i(\gamma_--\gamma_+)\sigma^y/2$. 
	It is straightforward to generalize Eq.~\eqref{eq:corrt_k_ferm} to the correlator 
	$\Gamma_{m,n}:=\langle \{r_m, r_n\}\rangle$~\cite{Carollo2022entdiss} (see Appendix~\ref{app:hc}).  
	From $\Gamma_{m,n}$  one can construct $\widetilde{\mathcal{C}}_{x,k,t}(t)$ in terms of 
	the $\eta_j$, which for bosons are eigenvectors of $\hat{\mathcal{L}}^\dagger$. 
	Here we consider the bosonic logarithmic negativity~\cite{vidal2002negativity}. 
	According to our conjecture, one has to compute the negativity 
	between the quasiparticles $e_n(x,k,t)$, which is~\cite{Audenaert2002negativity}  
	$e_n=-1/2\sum_{j}\ln
	\left[\min\left(1,\left|\mu_j\right|\right)\right]$. Here, $\mu_j$ are the
	eigenvalues of $i\Omega\Gamma^{T_2}$, with   
	\begin{equation}
	\label{eq:ptrans}
	\Gamma^{T_2}:=J_{\mathcal{A}_2} 
	\Gamma_{\mathcal{A}_1\cup\mathcal{A}_2} J_{\mathcal{A}_2}, \quad \Omega:= 
	(i\sigma^y)^{\oplus |\mathcal{A}_1\cup \mathcal{A}_2|}, 
	\end{equation}
	where $\Gamma_{{\mathcal{A}_1}\cup\mathcal{A}_2}(x,k,t)$ is  $\widetilde{\mathcal{C}}_{x,k,t}(t)$ 
	transformed into the $x_j(k):=(\eta_j(k)+\eta_j(k)^\dagger)/\sqrt{2}$ and 
	$p_j(k):=-i(\eta_j(k)-\eta^\dagger_j(k))/\sqrt{2}$ basis and restricted to quasiparticles in 
	$\mathcal{A}_1\cup \mathcal{A}_2$. In~\eqref{eq:ptrans}, $\Omega$ is  
	block-diagonal with equal blocks $i\sigma^y$. 
	$J_{\mathcal{A}_2}$ implements the partial transposition 
	with respect to $A_2$ on the correlation function $\Gamma$, and it is also block diagonal with the  
	$j$-th block equal to $\sigma_z$ if $j\in \mathcal{A}_2$ and to $\mathds{1}_2$ otherwise (see Appendix~\ref{app:chbasis} 	for a hydrodynamic ``derivation''). 
	For the mass quench, we have $n=2$~\cite{alba2018entanglement}. The negativity is 
	obtained from the eigenvalues of $J_{\mathcal{A}_2}\Gamma_{\mathcal{A}_1\cup\mathcal{A}_2} J_{\mathcal{A}_2}$. 
	They are (see Appendix~\ref{app:hc}) 
\begin{align}
\label{eq:lambda1}
& \lambda_1(k,t)=\frac{z_0^2}{z^2}\delta_t + \frac{z^4+1}{2g\; z^2}(1-\delta_t), \\
\label{eq:lambda2}
& \lambda_2(k,t)=\frac{z^2}{z_0^2}\delta_t + \frac{z^4+1}{2g\; z^2}(1-\delta_t) 
\end{align}
and $\lambda_3=-\lambda_1,\lambda_4=-\lambda_2$. The negativity $e_2$ is obtained as   
	\begin{multline}
	\label{eq:harmonic_neg}
	e_2(k,t)=-\ln\Big[\min\Big(1,\Big| \frac{\zeta_{k,m_0}}{\zeta_{k,m_1}}\delta_t + \frac{\zeta_{k,m_1}^2+1}{2g\; \zeta_{k,m_1}}(1-\delta_t) \Big|\Big) \Big]\\
	-\ln\Big[\min\Big(1,\Big|\frac{\zeta_{k,m_1}}{\zeta_{k,m_0}}\delta_t +\frac{\zeta_{k,m_1}^2+1}{2g\; \zeta_{k,m_1}}(1-\delta_t) \Big|\Big) \Big],
	\end{multline}
	where $\zeta_{k,m}:=[m^2 + 4\sin^2 (k/2)]^{1/2}$, 
	$\delta_t:=e^{-(\gamma_--\gamma_+)t}$, and 
	$g:=(\gamma_--\gamma_+)/(\gamma_-+\gamma_+)$. The derivation of~\eqref{eq:harmonic_neg} is cumbersome, 
	and it is reported in Appendix~\ref{app:hc}. 
	Interestingly, Eq.~\eqref{eq:harmonic_neg} implies that, except for $g=m_1 w=1$, 
	${\cal E}$ vanishes at  a finite value of $t/\ell$, which marks  
	the so-called \textit{sudden death of entanglement}~\cite{Yu_2009}. 
	Thus, at long times only the so-called 
	bound entanglement~\cite{horodecki1998mixed} is present.  
	The condition for the sudden death is derived analytically and involves only the slowest quasiparticles, 
	with $k\to0$, as they are the only ones entangling $A_1$ and $A_2$ at long times (see Appendix~\ref{app:sdeath}). 
	Fig.~\ref{fig:numdata} (f) 
	corroborates our conjecture (cf.~\eqref{eq:harmonic_neg} and~\eqref{eq:neg_leading}) 
	for the mass quench in the harmonic chain. 
	After the ``rise and fall'', the negativity 
	exhibits the sudden death, which is correctly described by the theory.

\section{Inhomogeneous initial states} 
	\label{sec:bipartite}

	\begin{figure}
		\includegraphics[width=.95\linewidth]{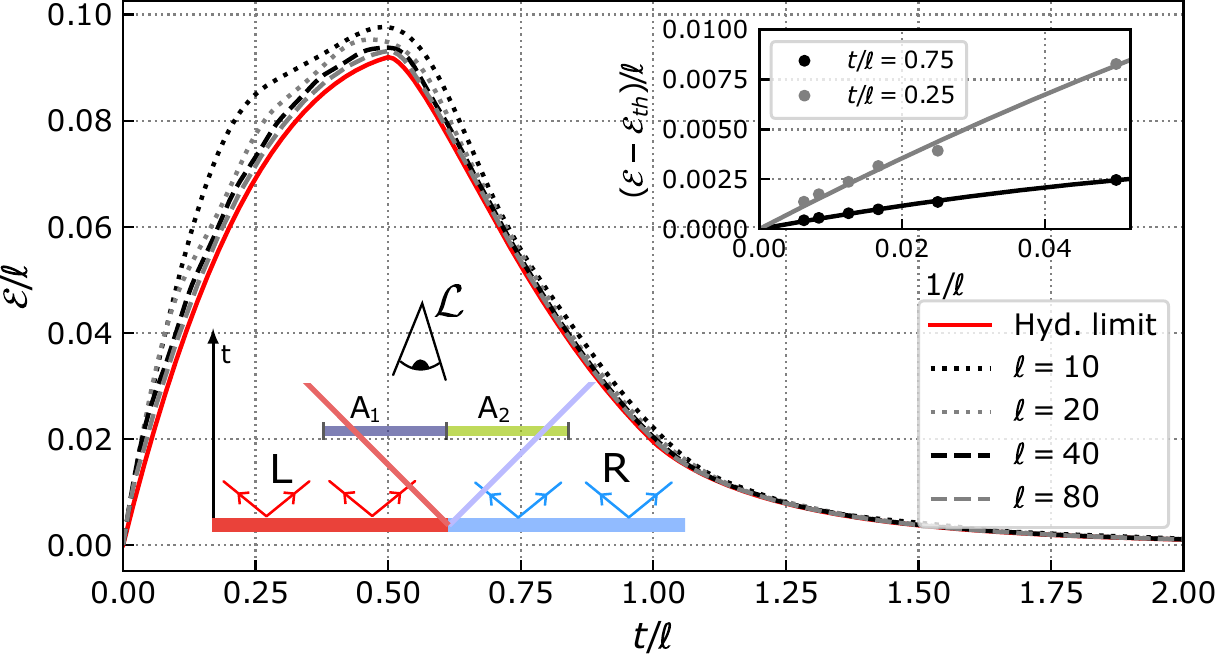}
		\caption{
			Negativity dynamics in the 
			tight-binding chain with gain and loss starting from an inhomogeneous state. 
			At $t=0$ the left and right parts of the chain are  prepared in the 
			dimer state and the N\'eel state, respectively. We plot $\cal{E}/\ell$ 
			for two adjacent equal-length intervals versus $t/\ell$. 
			The continuous curve is the analytic result in the  
			weakly-dissipative hydrodynamic limit. 
			The inset shows the finite-size corrections.  
		}
		\label{fig:ghd}
	\end{figure}
	We now consider the dynamics in the tight-binding chain with gain and losses starting 
	from the inhomogenous state $|L\rangle\otimes|R\rangle$, $|L\rangle$ and $|R\rangle$ being 
	the states of the two halves of the chain (see Fig.~\ref{fig:ghd}). 
	In the absence of dissipation the entanglement dynamics was 
	investigated already~\cite{bertini2018entanglementevolution,alba2018entanglementand,alba2019entanglement,alba2019towards}, 
	by using $GHD$~\cite{bertini-2016,olalla-2016}. 
	The mechanism for correlation spreading described in Fig.~\ref{fig:numdata} (c) 
	remains the same. However, now one has to ``transport'' the correlators 
	$\mathcal{C}_{x,k}^{\scriptscriptstyle(L)}$ and $\mathcal{C}_{x,k}^{\scriptscriptstyle(R)}$ 
	originating at $x<0$ and $x>0$, respectively. 
	Correspondingly, one obtains two contributions 
	$e_2^{\scriptscriptstyle(L)}$ and $e_2^{\scriptscriptstyle(R)}$. 
	If $A_1$ and $A_2$ are placed symmetrycally with respect to $x=0$ 
	(see Fig.~\ref{fig:ghd}), one obtains 
	\begin{equation}
	\label{eq:neg-aver}
	{\cal E}(t)=\ell\int_0^\pi \frac{dk}{2\pi} \Theta(k,t)\frac{e_{2}^{(L)}(k,t)+e_2^{(R)}(k,t)}{2}, 
	\end{equation}
	where $\Theta(k,t)$ is the same as in~\eqref{eq:neg_leading}. 
	Despite its simplicity, an \emph{ab initio} derivation 
	of~\eqref{eq:neg-aver} would be a daunting task. 
	
	In Fig.~\ref{fig:ghd} we consider the tight-binding chain and the quench from the state obtained by joining 
	the N\'eel state and the dimer state. 
	The continuous line in Fig.~\ref{fig:ghd} is Eq.~\eqref{eq:neg-aver} with 
	$e_2^{\scriptscriptstyle(L)}$ given by~\eqref{eq:neg_leading}. 
	For $e_2^{\scriptscriptstyle(R)}$ we used the result of Ref.~\cite{alba2022logarithmic}. 
	The agreement with~\eqref{eq:neg-aver} is perfect in the weakly-dissipative hydrodynamic limit. 
	

\section{Conclusions \& Outlook}
	\label{sec:concl}
	We derived the hydrodynamic framework for the 
	logarithmic negativity in generic bosonic and fermionic quadratic Lindblad systems. 
	Our results open new avenues for future research. 
	First and foremost, one could start an entanglement-based 
	classification of dissipative dynamics in generic free-fermion and free-boson systems 
	also  beyond one dimension, generalizing results at equilibrium~\cite{zhang2022criticality}. 
	Second, our approach is readily generalizable to inhomogeneous dissipation or 
	dynamics beyond the quench paradigm, such as Floquet driving~\cite{vancaspel2019dynamical}. 
	Third, it would be interesting to investigate 
	whether the hydrodynamic framework captures \emph{multipartite} entanglement~\cite{parez2024fate} between spatially-separated regions. 
	Clearly, it is of paramount importance to investigate the effect of interactions, going beyond quadratic systems. 
	Finally, Eq.~\eqref{eq:trace-hydro-diss} is not restricted to entanglement 
	measures, and it could be applied to the entanglement asymmetry~\cite{ares2023entanglement}, 
	generalizing the results of Ref.~\cite{caceffo2024entangled}. 

\section*{Acknowledgments} 

This study was carried out within the National Centre on HPC, Big Data and Quantum Computing - SPOKE 10 (Quantum Computing) and received funding from the European Union Next-GenerationEU - National Recovery and Resilience Plan (NRRP) – MISSION 4 COMPONENT 2, INVESTMENT N. 1.4 – CUP N. I53C22000690001. This work has been supported by the project “Artificially devised many-body quantum dynamics in low dimensions - ManyQLowD” funded by the MIUR Progetti di Ricerca di Rilevante Interesse Nazionale (PRIN) Bando 2022 - grant 2022R35ZBF.

	\bibliography{bibliography}

\appendix

\section{Effect of spatial symmetries and time reversal on quadratic operators}
\label{app:symm}

We briefly discuss some properties of the Fourier transform of block-circulant matrices 
that we employ in the following sections. Let our matrix $M$ have matrix elements 
$\lfloor M_{m,n}\rceil_{a,b}$, where the indices $m,n=1,...,L$ denote the position of the 
block and $a,b$ label the elements inside the block. Here we always consider 
fermionic and bosonic Hamiltonians with one-site translation invariance, meaning that 
$a,b=1,2$. 
Translation invariance implies that $M$ is block-circulant, allowing 
us to write
\begin{equation}
\label{eq:block_symbol}
\lfloor M_{m,n}\rceil=\lfloor M_{m-n}\rceil=\frac{1}{L}\sum_k \hat{m}_k e^{-ik(m-n)},
\end{equation}
where $k=2\pi j/L $ and $j=1,\dots,L$, 
and where  $\hat{m}_k$ is defined as the Fourier transform of $M$:
\begin{equation}
\hat{m}_k:=\sum_{m,n=1}^L\frac{1}{L} e^{ik(m-n)} \lfloor M_{m-n}\rceil.
\end{equation}
Let us now discuss how symmetries of $M$ are implemented in $\hat m_k$. It is straightforward to check that 
\begin{align}
\label{eq:symbol_herm}
& \lfloor M_{m-n}\rceil=\lfloor M_{n-m}\rceil^\dagger \quad  \Rightarrow \quad \hat{m}_k^\dagger=\hat{m}_k\\
\label{eq:symbol_sym}
& \lfloor M_{m-n}\rceil=\lfloor M_{n-m}\rceil^T \quad  \Rightarrow \quad \hat{m}_{-k}=\hat{m}_k^T\\
\label{eq:symbol_antisym}
& \lfloor M_{m-n}\rceil=-\lfloor M_{n-m}\rceil^T \quad \Rightarrow \quad \hat{m}_{-k}=-\hat{m}_k^T.
\end{align}
Furthermore, let us now consider the operator 
\begin{equation}
\mathcal{M}=\sum_{m,n,a,b}\lfloor M_{m,n}\rceil_{a,b}r_{m,a}r_{n,b}, 
\end{equation}
where $r_{m,a}$ are Majorana operators for fermionic models, or the conjugated operators $x_j,p_j$ for 
bosonic ones (see the main text). 
Let us consider the effect of the spatial inversion symmetry and of the 
time reversal symmetry on $\mathcal{M}$ and $M_{m,n}$. The spatial inversion $\hat R$ is implemented by 
a unitary operator acting on the local Majorana operators as 
\begin{equation}
\label{eq:spatial}
\hat{R}r_{m,a}\hat{R}=r_{-m,a}
\end{equation}
Thus, spatial inversion symmetry 
for the operator $\mathcal{M}$ implies that 
\begin{equation}
\label{eq:symbol_refl}
\lfloor M_{m-n}\rceil=\lfloor M_{n-m}\rceil \quad \Rightarrow \quad \hat{m}_{-k}=\hat{m}_k.
\end{equation}
We should notice there are several quadratic models that are \textit{not} symmetric 
under the spatial inversion as defined in~\eqref{eq:spatial}. A prominent example is the 
Kitaev chain, as it is easy to check.

Instead, the Kitaev chain is invariant under the time-reversal symmetry, which we now define. 
Time-inversion $\hat T$ is an \textit{anti-unitary} operator leaving the local bosonic and fermionic 
ladder operators  invariant. This implies that 
\begin{equation}
\hat{T}r_{j,o}\hat{T}=r_{j,o}, \quad 
\hat{T}r_{j,e}\hat{T}=-r_{j,e},
\end{equation}
where $r_{j,o}$ and $r_{j,e}$ are defined in the main text, and we used that $\hat Ti\hat T=-i$. 
Thus, invariance under time reversal implies that $M$ satisfies
\begin{equation}
\label{eq:symbol_trev}
\lfloor M_{m-n}\rceil=\sigma_z\lfloor M_{m-n}\rceil^* \sigma_z \quad \Rightarrow \quad \hat{m}_{-k}=\sigma_z\hat{m}_k^*\sigma_z.
\end{equation}
Here and in the following, $\sigma_{x,y,z}$ are the Pauli matrices.

In conclusion, the above results imply that if the  matrix $M$ is block-circulant, Hermitian, 
symmetric and invariant under spatial inversion, its Fourier transform is of the form 
\begin{equation}
\label{eq:first_app}
\hat{m}_k=\alpha \mathds{1} + \beta \sigma_x + \gamma \sigma_z,\; \alpha,\beta,\gamma \in \mathds{R}
\end{equation}
If $M$ is block-circulant, Hermitian, antisymmetric and invariant under spatial inversion, one has 
\begin{equation}
\label{eq:second}
\hat{m}_k=\alpha \sigma_y,\; \alpha \in \mathds{R}
\end{equation}
If $M$ is block-circulant, Hermitian, symmetric and time-reversal invariant, the Fourier transform satisfies
\begin{equation}
\label{eq:third}
\hat{m}_k=\alpha \mathds{1} + \beta \sigma_z,\; \alpha,\beta \in \mathds{R}.
\end{equation}
Finally, if $M$ is block-circulant, Hermitian, antisymmetric and invariant under time reversal, 
its $\hat m_k$ can be written as 
\begin{equation}
\label{eq:fourth}
\hat{m}_k=\alpha\sigma_x + \beta \sigma_y,\; \alpha,\beta \in \mathds{R}. 
\end{equation}
Clearly, Eq.~\eqref{eq:first_app} implies that $\hat m_k$ is real, which will turn out to 
be important to determine the quasiparticles for bosonic systems.
On the other hand, both~\eqref{eq:second} and~\eqref{eq:third} imply that $\mathrm{Tr}(\hat m_k)=0$, 
which will be  used  in Appendix~\ref{app:eigen} to determine the properties of fermionic quasiparticles.

\section{Effect of the dissipation on the propagating modes of the Liouvillian}
\label{app:eigen}
As discussed in the main text, a key ingredient in the hydrodynamic description 
of the negativity is the correct identification of the quasiparticles in the presence of 
dissipation. Let us remind that the state 
$\rho$ of the system evolves according to the master equation
\begin{multline}
\label{eq:lindblad_state}
\partial_t \rho = -i\left[H,\rho\right] \\
+ \sum_j\left(L_j \rho L_j^\dagger-\frac{1}{2} \left\{L_j^\dagger L_j,\rho\right\}\right):=-i \hat{\mathcal{L}} \rho,  
\end{multline}
where we have defined the Liouvillian superoperator $\hat{\mathcal{L}}=\hat{\mathcal{U}} + i \hat{\mathcal{D}}$. Here $\hat{\mathcal{U}}=\left[H,\cdot\right]$ the Hamiltonian part and $\hat{\mathcal{D}}$ 
is the dissipator, acting on the space $L(\mathcal{H})$ of linear operators on the (finite) 
Hilbert space $\mathcal{H}$ of the theory. $L(\mathcal{H})$ is a Hilbert space itself, with the scalar product being $\langle\langle A|B\rangle\rangle=\mathrm{Tr}\left[A^\dagger B \right]$. From Eq.~\eqref{eq:lindblad_state}, it is straightforward to derive the time evolution equation of the expected values of linear operators $\langle O \rangle=\mathrm{Tr}[\rho O]$ as:
\begin{equation}
\label{eq:lindblad_op}
\partial_t \langle O \rangle=i\langle\hat{\mathcal{L}}^\dagger O \rangle,
\end{equation}
with 
\begin{equation}
\hat{\mathcal{L}}^\dagger O= \left[H,O\right] -i \sum_j\left(L_j^\dagger O L_j-\frac{1}{2} \left\{L_j^\dagger L_j,O\right\}\right). 
\end{equation}
In the case of linear dissipation and quadratic Hamiltonian, we rewrite~\eqref{eq:lindblad_op}, 
in terms of local fermionic Majorana operators or bosonic position and momentum operators $x_j,p_j$. 
We obtain 
\begin{multline}
\label{eq:lindblad_lin}
\hat{\mathcal{L}}^\dagger O= \sum_{m,n=1}^{2L} h_{m,n}\left[r_m r_n,O\right]\\
-i\sum_{m,n=1}^{2L} K_{m,n} \left(r_m O r_n-\frac{1}{2}\left\{r_m r_n,O\right\}\right).
\end{multline}
The Kossakowski matrix $K$ is Hermitian and positive-semidefinite, 
and it can be decomposed as $K=A+iB$, with $A$ real symmetric and positive-semidefinite 
and $B$ real antisymmetric. For fermionic systems, the $r_j$ are the Majorana
fermions, whereas for bosons $r = (x_1, p_1, . . . , x_L, p_L)$ (see the main text). In the following, we will always assume invariance under one-site translations. 

\subsection{Fermionic systems}

Let us first consider the fermionic case, for which $h_{m,n}$ (cf.~\eqref{eq:lindblad_lin}) 
can be chosen antisymmetric and purely imaginary. Let us also assume that the Hamiltonian 
is invariant under either spatial inversion (Eq.~\eqref{eq:symbol_refl}) or time reversal 
(Eq.~\eqref{eq:symbol_trev}). In the former case the Fourier transform is $\hat{h}_k\propto\sigma_y$ (cf.~\eqref{eq:second}), while in 
the second one it is a real linear combination of $\sigma_x$ and $\sigma_y$ (cf.~\eqref{eq:fourth}). 

For unitary dynamics, we have $\hat{\mathcal{D}}=0$. Moreover, the Liouvillian superoperator $\hat {\mathcal{L}}$ 
maps local Majorana operators to linear combinations of local Majorana operators. 
Thus, it is natural to construct the  single-particle propagating modes as eigenvectors of 
the Liouvillian that are linear combinations of  
Majorana operators. The eigenvector $\eta^\dagger$ of the adjoint Liouvillian will satisfy 
$\left[H,\eta^\dagger\right]=\lambda \eta^\dagger$. 
Let us now exploit translation invariance. In the thermodynamic limit $L\to\infty$ the  
Hamiltonian is rewritten as
\begin{equation}
\label{eq:ham_k_ferm}
H=\int_{-\pi}^\pi\frac{dk}{2\pi} \begin{pmatrix}
r_o(-k)&r_e(-k)
\end{pmatrix} \hat{h}_{k}\begin{pmatrix}
r_o(k)\\r_e(k)
\end{pmatrix},
\end{equation}
where $r_o(k):= \frac{1}{\sqrt{L}}\sum_{j=1}^{L}e^{ikj}r_{2j-1}=c_k+c_{-k}^\dagger$ 
and $r_e(k) := \frac{1}{\sqrt{L}}\sum_{j=1}^{L}e^{ikj}r_{2j}=
i(c_k-c_{-k}^\dagger)$ are the Fourier transforms of odd and even Majorana fermions, respectively.
It is then apparent that for each quasimomentum $k$, the $2\times2$ subspace generated 
by the operators $r_o(k)$ and $r_e(k)$ is left invariant by the dynamics. Indeed, one has 
\begin{equation}
\label{eq:eigen_unitary_2}
\begin{pmatrix}
[H,r_o(k)] \\ [H,r_e(k)]
\end{pmatrix}=-4\hat{h}_k\begin{pmatrix}
r_o(k)\\r_e(k)
\end{pmatrix},
\end{equation}
where we have used Eq.~\eqref{eq:symbol_antisym}. From~\eqref{eq:eigen_unitary_2} it follows 
that the propagating modes are linear combinations of  
$r_o(k)$ and $r_e(k)$. The coefficients $w(k)$ of the combination satisfy
\begin{equation}
\label{eq:eigen_unitary_3}
w^T(k)(-4\hat{h}_k)=\lambda w^T(k).
\end{equation}
If $w_\pm(k)$ are the two orthonormal left eigenvectors of $4\hat{h}_k$, 
their eigenvalues are $\lambda_\pm=\pm \varepsilon(k)$,  
since $\mathrm{Tr}[\hat{h}_k]=0$, and they are real.  We can  
define $\eta_k,\eta_{-k}$ as
\begin{equation}
\label{eq:eigen_unitary_4}
\begin{pmatrix}
\eta_k\\\eta_{-k}^\dagger
\end{pmatrix}= \frac{1}{\sqrt{2}} V_k \begin{pmatrix}
r_o(k)\\r_e(k)
\end{pmatrix},\quad V_k^T=\{w_+,w_-\}, 
\end{equation}
i.e., the unitary matrix $V_k$ has rows $w^T_+(k)$, $w^T_-(k)$. 
Here $k\in [0,\pi]$ and the labels $k, -k$ in the left-hand side in~\eqref{eq:eigen_unitary_4} 
are consistent with the right-hand side.  Indeed, 
by using~\eqref{eq:symbol_herm} and~\eqref{eq:symbol_antisym} in~\eqref{eq:eigen_unitary_3} one obtains that 
\begin{equation}
w^\dagger (-4\hat{h}_{-k})=-\lambda w^\dagger. 
\end{equation}
Thus, $w_\pm(-k)=w_\mp(k)^*$, implying that the rows 
of~\eqref{eq:eigen_unitary_4} for $-k$ are the adjoints of those for $k$.

We can then naturally define two families of quasiparticles as $\eta_1^\dagger(k)=\eta_k^\dagger$, 
and $\eta^\dagger_2(k)=\eta^\dagger_{-k}$, with associated eigenvalues 
$\varepsilon_1(k)=\varepsilon_2(k)=\varepsilon(k)$. The anticommutation relation 
$\{\eta^\dagger_i(k),\eta_j(h)\}=\delta_{i,j}\delta_{k,h}$ is fulfilled because of the unitarity of $V_k$. 
In the thermodynamic limit $L\to \infty$, the Hamiltonian can be rewritten in terms of $\eta_1,\eta_2$ as
\begin{equation}
\label{eq:eigen_unitary_ham}
H=\int_{0}^\pi \frac{dk}{2\pi} \varepsilon(k) \left(\eta_1(k)^\dagger \eta_1(k)+\eta_2(k)^\dagger \eta_2(k) -1\right).
\end{equation}
The propagation velocity of a species of quasiparticles is the group velocity 
obtained from the dispersion $\epsilon_j(k)$ in front of $\eta_j(k)^\dagger \eta_j(k)$ 
after the Hamiltonian has been reduced in a form like~\eqref{eq:eigen_unitary_ham}. 
An important caveat is that one has to consider the  dispersion of
the excitations with respect to the \textit{physical} quasimomentum $k\in[-\pi,\pi]$ and not the 
folded one~\cite{caceffo2024entangled} $k\in[0,\pi]$, thus $v_1(k)=\varepsilon'(k)$, 
and $v_2(k)=- \varepsilon'(k)=-v_1(k)$. 
Finally, we observe that  the condition $\mathrm{Tr} [\hat{h}_k]=0$, which is a consequence either 
of time reversal or of spatial inversion symmetry, ensures $v_1(k)=-v_2(k)$.

Let us now discuss the dissipative case $\hat{\mathcal{D}}\neq 0$ in~\eqref{eq:lindblad_op}. As a further assumption, we require the dissipator to have the same symmetry as the Hamiltonian, that is to say, the Kossakowski matrix $K$ (cf.~\eqref{eq:lindblad_op}) satisfies either~\eqref{eq:symbol_refl} or~\eqref{eq:symbol_trev}. First, 
the subspace of operators that are linear in the fermionic local operators is not left 
invariant by the adjoint Liouvillian $\hat{\mathcal{L}}^\dagger$. It is 
easy to check that the dissipator $\hat{\mathcal{D}}$ applied to linear combinations of fermionic operators  
produces also cubic operators. Thus, following~\cite{Carollo2022entdiss,starchl2022relaxation}, 
we consider \textit{quadratic} operators of the form $\chi=\eta^\dagger \eta$, where 
$\eta$ is linear in fermion operators. We require that $\chi$ satisfies the equation
\begin{equation}
\label{eq:eigen_diss_ferm1}
\hat{\mathcal{L}}^\dagger \chi= \xi (\chi + c), 
\end{equation}
with $\xi$ and $c$ to be determined. Thus, we identify $\eta^\dagger$ as 
the quasiparticle. This means that $\eta^\dagger$ is a quasiparticle 
if its density $\chi=\eta^\dagger\eta$ evolves according to~\eqref{eq:eigen_diss_ferm1}. 
In the following we will show that the associated eigenvalue $\xi$ is 
purely imaginary. Notice that  a real part would give an oscillating behaviour which is not expected for the 
density (in analogy with the unitary case). Moreover,  $\xi$ will have a  
positive imaginary part, reflecting that the quasiparticles  decay. 

Let us first show that in a one-site translation invariant system there are always 
two independent species of quasiparticles (in the sense of~\eqref{eq:eigen_diss_ferm1}), 
$\eta_k^\dagger$, $\eta_{-k}^\dagger$ with $k\in[0,\pi]$.  
We anticipate that the $\eta_k$ and $\eta_{-k}$  satisfy canonical anticommutation relations only 
in the limit of weak dissipation. Moreover,  
the quadratic operator $\chi'_k=\left[\eta_k,\eta_{-k}\right]$ will also be an eigenvector 
of the adjoint Liouvillian as 
\begin{equation}
\label{eq:eigen_diss_ferm2}
\hat{\mathcal{L}}^\dagger \chi'_k= \xi'_k (\chi'_k + c'_k). 
\end{equation}
In analogy with the unitary case, the real part of $\xi'$, that 
is responsible for the oscillatory behaviour of $\chi'$, will 
provide the dispersion relation of the quasiparticles $\eta_j^\dagger$. Thus, 
$\xi'$ is needed to identify the propagation velocities of the quasiparticles. 

To proceed, let us write $\chi=\sum_{m,n=1}^L Q_{m,n}r_m r_n$, where  $Q$ 
can be chosen Hermitian and antisymmetric, since $\chi$ is self-adjoint. Then, 
the eigenvalue equation~\eqref{eq:eigen_diss_ferm1} becomes
\begin{multline}
\label{eq:chi}
\hat{\mathcal{L}}^\dagger \chi=\sum_{m,n=1}^{2L}\left[\left(4h+2iA\right)Q + Q\left(-4h+2iA\right)
\right]_{m,n} r_m r_m \\
-4\mathrm{Tr}\left[BQ\right] =\xi \left(\chi+c\right),
\end{multline}
where we used the decomposition of the Kossakowski matrix as 
$K=A+iB$, and we exploited the symmetries $h^T=-h$, $A^T=A$, 
$Q^T=-Q$. Moreover, if we assume translation invariance, Eq.~\eqref{eq:chi}  
becomes the eigenvalue equation for the $2\times 2$ Fourier transform 
$\hat{q}_{k'}$ of $Q$ as 
\begin{equation}
\label{eq:chi_trinv}
\left(4\hat{h}_{k'} +2i\hat{A}_{k'}\right)\hat{q}_{k'}- \hat{q}_{k'} \left(4\hat{h}_{k'}+2i\hat{A}_{k'}\right)^\dagger =
\xi \hat{q}_{k'},\quad\forall k'.
\end{equation}
Now, let us consider the case of ``small'' dissipation (i.e., small $\hat{A}_k$) in which the 
solution of~\eqref{eq:chi_trinv} is simpler. Notice that this is not a very restrictive assumption, since
our main interest is the limit of vanishing dissipation.  
Then, the matrix $\hat{X}_k:=4\hat{h}_k +2i\hat{A}_k$ is always diagonalizable. 
The symmetries of $h$ and $A$ (Eqs.~\eqref{eq:first_app},~\eqref{eq:second} 
or~\eqref{eq:third},~\eqref{eq:fourth}) imply that $\hat{x}_k$ has two distinct eigenvalues 
of the form $\lambda_k=\pm \omega_k +i \gamma_k$, 
where both $\omega_k$ and $\gamma_k$ are real. We choose $\omega_k>0$ and we have 
$\gamma_k=\mathrm{Tr}[\hat{A}_k]>0$, as expected. 
This means that the density of the quasiparticles decays with time, as anticipated.  If $w_\pm(k)$ 
are the associated (column) eigenvectors of $\hat x_k$, we can construct a solution of~\eqref{eq:chi_trinv} 
as $\hat{q}^{\scriptscriptstyle(k)}_{\pm,k'}=
w_\pm(k)w_\pm^\dagger(k)\delta_{k,k'}$ satisfying Eq.~\eqref{eq:chi_trinv}, with 
$\xi=\xi^\pm_k:=2i\gamma_k$. Thus, if the matrix $V_k$ with columns $w_+(k), w_-(k)$ 
diagonalizes $\hat x_k$ as $\hat x_k=V_k D_k V_k^{-1}$, we can define the quasiparticles as
\begin{equation}
\label{eq:eigen_diss_ferm3}
\begin{pmatrix}
\eta_k\\\eta_{-k}^\dagger
\end{pmatrix}=\frac{1}{\sqrt{2}}V_k^\dagger \begin{pmatrix}
r_o(k)\\r_e(k)
\end{pmatrix},
\end{equation}
where the factor $1/\sqrt{2}$ is a normalization we will need 
later. Notice that in the unitary case $\hat{A}_k=0$, and we recover 
Eq.~\eqref{eq:eigen_unitary_4}, as it is clear from the 
definitions of $w_\pm(k)$. 

We now define two families of quasiparticles, in analogy with the unitary 
case, as $\eta_1(k)^\dagger=\eta_k^\dagger$, 
$\eta_2(k)^\dagger=\eta^\dagger_{-k}$. They satisfy anticommutation relations
\begin{equation}
\label{eq:eigen_diss_anticomm}
\begin{pmatrix}
\left\{\eta_1(k),\eta_1(k')^\dagger\right\}&\left\{\eta_1(k), \eta_2(k')\right\}\\
\left\{\eta_2(k)^\dagger, \eta_1(k')^\dagger\right\}&\left\{\eta_2(k)^\dagger, \eta_2(k')\right\}
\end{pmatrix}= \delta_{k,k'} V_k^\dagger V_k,
\end{equation}
that in the limit of small dissipation become the canonical anticommutation 
relations, since the matrix $V_k$ diagonalizing $\hat x_k$ becomes unitary. 

Finally, we check that the operator $\chi'_k=\left[\eta_k,\eta_{-k}\right]=\left[\eta_1(k),
\eta_2(k)\right]$ is also an eigenvector of $\hat{\mathcal{L}}$ in the 
sense of~\eqref{eq:eigen_diss_ferm2}. Indeed, we can write it as 
\begin{equation}
\chi'_k=\sum_{m,n=1}^L P^{\scriptscriptstyle(k)}_{m,n}r_m r_n,
\end{equation}
where $P^{\scriptscriptstyle(k)}$ is  block circulant and has $2\times 2$ Fourier transform 
\begin{equation}
\hat{p}^{\scriptscriptstyle(k)}_{k'}=\frac{1}{2}
\left(\delta_{k,-k'}w^*_+(k)w^T_-(k)-\delta_{k,k'}w_-(k)w^\dagger_+(k) \right). 
\end{equation}
Now, the eigenvalue equation~\eqref{eq:eigen_diss_ferm2} for $\chi'_k$ becomes
\begin{multline}
\label{eq:eigen_diss_ferm4}
\hat{\mathcal{L}} \chi'_k=\xi_k'(\chi'_k+c'_k)=
\sum_{m,n=1}^{2L}\Big(4\left[h,P^{(k)}\right] \\+i \left\{A,P^{(k)}-\left(P^{(k)}
\right)^T\right\}\Big)_{m,n} r_m r_n
-4\mathrm{Tr}\left[BP^{(k)}\right]
\end{multline}
which implies, exploiting the fact that $\left(\hat{p}^{(k)}_{-k'}\right)^T
=-\hat{p}^{(k)}_{k'}$, that $\hat{p}^{(k)}_{k'}$ must satisfy:
\begin{equation}
\label{eq:eigen_diss_ferm5}
\hat{x}_{k'}  \hat{p}^{(k)}_{k'} - \hat{p}^{(k)}_{k'} \hat{x}_{k'}^\dagger=\xi_k' \hat{p}^{(k)}_{k'},
\end{equation}
which is true for $\xi_k'=-2\omega_k+2i\gamma_k$. In the unitary case, i.e., without dissipation, 
one has  $\hat{\mathcal{L}}\chi'_k=-2\varepsilon(k)\chi'_k$, with $\varepsilon(k)$ the dispersion of the 
model. This means that the eigenvalue equation for $\chi'$ gives 
access to the propagation velocities of the two quasiparticles  as 
$v_1(k)=-v_2(k)=\partial_k\omega_k$. 
A similar idea holds in the presence of dissipation~\cite{starchl2022relaxation}, where the real part of 
$\xi'_k$ gives access to the velocities of the quasiparticles. 
Moreover, in the limit of weak dissipation~\cite{Carollo2022entdiss} the velocities of the  quasiparticles 
become the same as in the unitary case, i.e., $\pm\varepsilon'(k)$. 
Incidentally, we notice that the symmetry under spatial reflection and time-reversal was crucial to 
derive the results above. Indeed, dropping the symmetry requirements on the 
dissipation would produce different dissipation rates for the densities of the 
two quasiparticles of the pair, distinguishing between right and left movers. 
Indeed, $x_k$ would have eigenvalues $\lambda_k=\pm \omega_k \pm i\delta_k 
+i \gamma_k$, with $\delta_k \in \mathds{R}$, implying $\xi_k^\pm=2i
\left(\gamma_k\pm\delta_k\right)$ (it is quite cumbersome but straightforward to 
prove that the imaginary part of $\xi_k^\pm$ would still be positive). For 
the time-reversal symmetric Kitaev chain, the dissipations 
without time-reversal symmetry turn out to be the ``odd dissipations" of~\cite{Alba2022entdiss}. 

\subsection{Bosonic systems}

Let us now consider the bosonic case, for which the matrix $h$ in 
Eq.~\eqref{eq:lindblad_lin} can be always chosen real and symmetric, 
and is usually positive-definite. We also assume invariance 
under spatial inversion (Eq.~\eqref{eq:symbol_refl}) for both the 
Hamiltonian and the dissipation. The bosonic  case is easier to deal 
with than the fermionic one, since the subspace of linear operators 
is always invariant under the action of $\hat{\mathcal{L}}^\dagger$.  
Thus, we can construct the quasiparticles  as 
	linear combinations of $r_j$. If we write  
$\eta=\sum_j w_j r_j$, the vector $w$ must satisfy the equation 
\begin{equation}
\label{eq:eigen_diss_bos1}
i\left(2 h - B\right) \Omega w=\lambda w, 
\end{equation}
with $\lambda$ the eigenvalue, and the Hamiltonian matrix $h$ and $B$ 
as in~\eqref{eq:lindblad_lin}. The symplectic form $\Omega$ encodes the commutation relations between the bosonic operators and reads $\lfloor \Omega_{m,n}\rceil=i\sigma_y\delta_{m,n}$. 
Since we have translation invariance, we can look for eigenvectors belonging 
to the $2\times 2$ invariant subspace generated by $x_k:=\frac{1}{\sqrt{L}}\sum_{j=1}^Le^{ikj}x_j$ and $p_k:=\frac{1}{\sqrt{L}}\sum_{j=1}^Le^{ikj}p_j$, which are 
the Fourier transforms of position and momentum operators. In that subspace, the eigenvalue problem 
in~\eqref{eq:eigen_diss_bos1} becomes 
\begin{equation}
\label{eq:eigen_diss_bos2}
i\left(2\hat{h}_{-k}-\hat{B}_{-k}\right)(i \sigma_y) w_k =\lambda_k w_k.
\end{equation}
The invariance under spatial inversion, as we have shown in Appendix~\ref{app:symm} above, 
implies that $i\hat{B}_k=b_k\sigma_y, \; b_k\in \mathds{R}$. In turn, this  
implies $-i\hat{B}_{-k}(i\sigma_y) =-ib_{-k}\mathds{1}_{2\times 2}$. In conclusion, spatial 
inversion symmetry ensures that the quasiparticles are the same quasiparticles of the  
Hamiltonian. Since $h$ is positive-definite, $\hat{h}_{-k}$ is also positive. Then, 
for Williamson's theorem it always satisfies
\begin{multline}
\label{eq:sympl_eigs}
i\hat{h}_{-k} (i \sigma_y) =s_k^{-1} \begin{pmatrix}
0&i \omega_k\\-i\omega_k &0
\end{pmatrix} s_k=\\
\frac{1}{2}s_k^{-1}\begin{pmatrix}
1&1\\i&-i
\end{pmatrix}\begin{pmatrix}
-\omega_k&0\\0&\omega_k
\end{pmatrix}\begin{pmatrix}
1&-i\\1&i
\end{pmatrix}s_k,
\end{multline}
where the real matrix $s_k$ is symplectic and satisfies $s_k \hat{h}_{-k}s_k^T=
\text{diag}(\omega_k,\omega_k)$, $\omega_k \in \mathds{R}^+$~\cite{Arnold}. Thus, we can 
define the quasiparticles as
\begin{equation}
\label{eq:eigen_diss_bos3}
\begin{pmatrix}
\eta_k\\\eta_{-k}^\dagger
\end{pmatrix}=\frac{1}{\sqrt{2}}\begin{pmatrix}
1&i\\1&-i
\end{pmatrix}\left(s_k^{-1}\right)^T\begin{pmatrix}
x_k\\p_k
\end{pmatrix},
\end{equation}
where the rows of the matrix are the transposed of the two eigenvectors 
$w_\mp(k)$ satisfying~\eqref{eq:eigen_diss_bos2}, with eigenvalues 
$\mp2\omega_k-ib_{-k}$. As in the fermionic case, $k\in[0,\pi]$, since 
in Eq.~\eqref{eq:eigen_diss_bos3} for $-k$ we have the adjoint operators 
$\eta_{-k}$, $\eta_k^\dagger$. This follows from the fact that $s_{-k}=s_k$, because  
the spatial inversion symmetry implies $\hat{h}_{-k}=\hat{h}_k$. 
We identify our two families of quasiparticles as $\eta^\dagger_1(k)=\eta_k^\dagger$, 
$\eta^\dagger_2(k)=\eta_{-k}^\dagger$. The fact that $s_k$ is symplectic ensures that quasiparticles 
have the correct commutation relations. In terms of these two families, the Hamiltonian can be rewritten (in the thermodynamic limit $L\to \infty$) as
\begin{equation}
H=\int_{0}^\pi \frac{dk}{2\pi} 2\omega_k (\eta_1(k)^\dagger \eta_1(k)+\eta_2(k)^\dagger \eta_2(k)+1).
\end{equation}
The velocities of the two families of quasiparticles are given by the group velocity 
obtained from their energy dispersion, with the same caveat as in the 
fermionic case that they have to be extracted from the dispersion before the folding. One has 
$v_1(k)=-v_2(k)=2\partial_k\omega_k$.

\subsection{Summary and generalization}

The construction above shows that for fermions the effect of the dissipation 
is simply a mixing between the quasiparticles $\eta_1(k)$, $\eta_2(k)^\dagger$ of the Hamiltonian, 
as it is clear by comparing Eqs.~\eqref{eq:eigen_unitary_4} and~\eqref{eq:eigen_diss_ferm3}. 
Thus, if the initial state is translationally invariant, the two-point 
function is block-diagonal in the quasimomentum space with respect to 
the Hamiltonian quasiparticles. The time-dependent correlator 
$\widetilde{\mathcal{C}}_{x,k,t}(t)$ can be reconstructed by solving the time evolution 
equation of the Fourier transform $\hat{g}_k(t)$ of the two-point function, 
that contains all the non-zero correlations allowed by the translational symmetry.

For bosons, typical dissipations  have no 
effect on the structure of the quasiparticles of the Hamiltonian. 
A case of dissipation that does not respect 
spatial inversion symmetry, implying that the quasiparticles are modified by 
the dissipation, was studied in~\cite{Carollo2022entdiss}. 
In the setup of Ref.~\cite{Carollo2022entdiss} the net effect of the dissipation 
is, again, a mixing between $\eta_1(k)^\dagger$ and $\eta_2(k)$. 
Thus, also in the bosonic case if the initial state is translationally invariant, 
the two point function is block-diagonal in the basis $\eta_1,\eta_2$, 
and the time dependence of $\widetilde{\mathcal{C}}_{x,k,t}(t)$ can be reconstructed from the Fourier transform of the two-point function $\hat{g}(k,t)$. 

When the translational symmetry is weakened (for example, becomes a two-site translation invariance), 
to preserve the block-diagonality one has to write the two-point function in quasimomentum space 
in terms of larger number $n$ of families of 
quasiparticles. In the unitary case, this is equivalent to require that we can 
rewrite the Hamiltonian in the thermodynamic limit $L\to \infty$ as~\cite{caceffo2024entangled}
\begin{equation}
\label{eq:ham_multi}
H=\int_0^{\frac{2\pi}{n}}\frac{dk}{2\pi} \sum_{j=1}^n \epsilon_j(k) \eta_j(k)^\dagger \eta_j(k) + C,
\end{equation}
where the quasiparticles $\eta_j$ satisfy canonical commutation or anticommutation 
relations, and the velocities of the species are identified as 
$v_j(k)=\pm \epsilon_j'(k)$, where the sign, as discussed above, is the one obtained 
from the energy dispersion before folding the quasimomentum. 
We expect that the effect on this multiplet structure of a dissipation with at least 
the same spatial symmetry as the Hamiltonian should be simply to mix the $\eta_i, \eta_j^\dagger$ 
with fixed quasimomentum $k$. Thus, one could expect that the procedure to 
construct the quasiparticles we presented can be generalized to arbitrary multiplets. 

\section{Hydrodynamic derivation of Eq.~\eqref{eq:trace-hydro} and Eq.~\eqref{eq:trace-hydro-diss}}
\label{app:chbasis}

In this Appendix, we give a more detailed justification of our conjectures~\eqref{eq:trace-hydro} 
and~\eqref{eq:trace-hydro-diss} in the main text. We show how they rely on the fact that quasiparticles  obey 
canonical (anti)commutation relations, and on the hydrodynamic assumptions of independent 
mesoscopic cells and ballistic propagation of the correlations (cf. Fig.~\ref{fig:numdata} (c)). 
We also show that, for bosons, the way we construct the position-momentum correlator $\Gamma(x,k,t)$ 
associated to a multiplet (see Eq.~\eqref{eq:ptrans}) 
is the correct one, even though it does not correspond to the correlator of any set of Fourier-transformed position-momentum operators. In the following, it will be useful 
to employ the notation $m^{\oplus}$ to indicate a block-diagonal matrix with the proper number (depending on the context) of equal blocks $m$ on the diagonal. 
Moreover, with $m\sim n$ we mean that the matrices $m$ and $n$ are similar (thus, they have the same spectrum).

Let us consider any local observable $f(\rho_A(t))$ that on Gaussian states can be evaluated as $f(\rho_A(t))=\mathrm{Tr}\left[\mathcal{F}\left(C_A(t)\right)\right]$, where $C_A(t)$ is the two-point function in real space restricted to a subsystem $A$ and $\mathcal{F}$ a suitable function related to $f$.
We assume that in the hydrodynamic limit we can instead calculate 
the trace of the same function $\mathcal{F}$ of the correlator ${\mathcal C}_{t}=\bigoplus_{x,k}\widetilde{\mathcal C}_{x,k,t}(t)$, where 
the restriction on a subsystem $A$ becomes the restriction on the quasiparticles in that subsystem. 
Thus, in our conjecture, we assume that in the hydrodynamic limit
$\mathcal{F}(C_A(t))$ has the same spectrum as $\mathcal{F}({\mathcal C}_{t}^{\scriptscriptstyle(A)})$. Provided that the quasiparticles 
satisfy canonical (anti)commutation relations, this is simply a consequence of the fact that the mesoscopic cells are independent and 
the dynamics in each cell produces the same quasiparticles of the full chain. This is justified because the cell is ``large''. 
Indeed, within  this approximation, the transformation between $C_A(t)$ and ${\mathcal C}_{t}^{\scriptscriptstyle(A)}$ is of the form
\begin{equation}
\label{eq:chbasis}
C_A(t)=u^\oplus{\mathcal C}_{t}^{(A)}(u^\dagger)^\oplus,
\end{equation}
where each copy of $u$ acts on a cell contained in $A$.
Now, it is useful to distinguish between the fermionic case and the bosonic one. The fermionic case 
is much easier. Indeed, if the quasiparticles obey canonical anticommutation relations, 
then $u$ is unitary and $C_A(t)\sim{\mathcal C}_{t}^{(A)}$ for any subsystem $A$, thus one simply has to check that the required $\mathcal{F}$ is well-behaved, i.e., it preserves the similarity.
In this paper, we are interested in the case of the fermionic negativity. One has~\cite{Eisert_2018neg}
\begin{equation}
\label{eq:fermneg}
\mathcal{E}(t)=\frac{1}{2}\left[S_{1/2}(\Gamma_\times(t))-S_2(\Gamma_{A_1\cup A_2}(t))\right],
\end{equation}
where we define
\begin{equation}
\label{eq:gammax}
\Gamma_\times:=[(\mathds{1}+\Gamma_{A_1\cup A_2}^2)/2]^{-1} 
\mathrm{diag}(\Gamma_{A_1},-\Gamma_{A_2})
\end{equation}
and $\Gamma_A:=2\mathcal{C}_A(t)-\mathds{1}$. In~\eqref{eq:fermneg}, we also defined 
\begin{equation}
	S_\alpha=\frac{1}{2(1-\alpha)}\ln\left[\left(\frac{1+x}{2}\right)^\alpha + \left(\frac{1-x}{2}\right)^\alpha \right]. 
\end{equation}
$S_\alpha(\Gamma_A)$ is the R\'enyi-$\alpha$ entropy of a subsystem $A$~\cite{peschel2009reduced}. With these definitions, 
it is apparent that $\mathcal{F}(C_A(t))\sim\mathcal{F}(\mathcal{C}_{t}^{(A)})$. We thus obtain Eq.~\eqref{eq:neg-def} 
in the main text. 
Notice that the generalization to other quantities, such as the charged moments of the entanglement asymmetry~\cite{caceffo2024entangled}, 
is straightforward. An important caveat is that in the presence of dissipation the quasiparticles 
might not satisfy the canonical anticommutation relations for finite dissipation rates, 
in contrast with the weak dissipation limit (see Eq.~\eqref{eq:eigen_diss_anticomm}). This means that for 
finite dissipation Eq.~\eqref{eq:trace-hydro-diss}  has to be modified. Precisely, Eq.~\eqref{eq:trace-hydro-diss} will depend 
on the matrix $V_k$ in Eq.~\eqref{eq:eigen_diss_anticomm}. This strategy has been employed 
in~\cite{starchl2022relaxation} for the von Neumann entropy.

In the bosonic case, the preservation of the commutation relations implies (notice that we are working in the ladder operators' basis)
\begin{equation}
\label{eq:commpres}
u \sigma_z^\oplus u^\dagger=\sigma_z^\oplus=u^\dagger \sigma_z^\oplus u,
\end{equation}
where the second equality follows from the fact that $\sigma_z^{-1}=\sigma_z$. Then, $C_A(t)$ 
and ${\mathcal C}_{t}^{(A)}$ have different spectra. This is not very surprising, since for bosonic correlators the relevant spectrum
is usually the \textit{symplectic} one. Indeed, the issue can be solved by the form of $\mathcal{F}$, which can ensure that $\mathcal{F}(C_A(t))\sim\mathcal{F}(\mathcal{C}_{t}^{(A)})$.
This is exactly what happens for our case of interest, the bosonic logarithmic negativity, as we are going to show.
For bosons the negativity is obtained as~\cite{Audenaert2002negativity}
\begin{equation}
\label{eq:bosneg}
\mathcal{E}(t)=-1/2\sum_{j}\ln\left[\min\left(1,\left|\mu_j\right|\right)\right],
\end{equation}
where $\mu_j(t)$ are the eigenvalues of $i \Omega J_{A_2} \Gamma_{A}(t)  J_{A_2}$, with $\Gamma_{A}$ the two-point function 
restricted to $A:=A_1\cup A_2$, $\Omega=i\sigma_y^\oplus$, and $J_{A_2}$ is 
a matrix that represents the effect of the partial transposition of the state with respect to $A_2$ on 
$\Gamma_{A}$, reading $J_{A_2}=\mathds{1}_2^{\oplus|A_1|} \oplus \sigma_z^{\oplus|A_2|}$, thus acting 
as $\mathds{1}$ on each site of $A_1$ and as $\sigma_z$ on each site of $A_2$. Notice that Eq.~\eqref{eq:bosneg} does not provide explicitly the form of $\mathcal{F}$ for the bosonic negativity. Now,
the change of basis from the position and momentum $x_j$, $p_j$ to the bosonic ladder operators $b_j$ is 
implemented by a unitary matrix $R$ as 
\begin{equation}
\label{eq:r}
\begin{pmatrix}
x_j\\p_j
\end{pmatrix}=\frac{1}{\sqrt{2}}\begin{pmatrix}
1&1\\i&-i
\end{pmatrix}\begin{pmatrix}
b_j^\dagger\\b_j
\end{pmatrix}:=R\begin{pmatrix}
b_j^\dagger\\b_j
\end{pmatrix}.
\end{equation}
Next, let us observe that (with $J_{A_2}$ from now on called simply $J$ for brevity)
\begin{equation}
\label{eq:can_to_lad}
i\Omega J\Gamma_A (t) J =- R^{\oplus} \sigma_z^{\oplus } J' C_A(t) J' (R^\dagger)^{\oplus}, 
\end{equation}
with 
\begin{equation}
J':=\mathds{1}_2^{\oplus }\oplus\sigma_x^{\oplus}.
\end{equation}
Eq.~\eqref{eq:can_to_lad} implies that  
\begin{equation} 
	\label{eq:first}
i\Omega J\Gamma_A(t) J\sim -\sigma_z^{\oplus}J'C_A(t)J'=-J''\sigma_z^{\oplus}C_A(t) J',
\end{equation}
with 
\begin{equation}
J'':=\mathds{1}_2^{\oplus}
\oplus(-\sigma_x)^{\oplus}, 
\end{equation}
providing the definition of the bosonic negativity as $\mathrm{Tr}\mathcal{F}(C_A):=-\mathrm{Tr}(J''\sigma_z^\oplus C_A J')$. 
We now apply~\eqref{eq:chbasis} and~\eqref{eq:commpres}, obtaining 
\begin{multline}
-J''\sigma_z^{\oplus}C_A(t) J'=-J''\sigma_z^{\oplus}u^\oplus{\mathcal C}_{t}^{(A)}(u^\dagger)^\oplus J'=\\
-J''(u^{-1})^{\dagger \oplus}\sigma_z^{\oplus}{\mathcal C}_{t}^{(A)}(u^\dagger)^\oplus J'.
\end{multline}
Then, since $J'^2=J''^2=\mathds{1}$, we have 
\begin{multline}
	\label{eq:last}
-J''\sigma_z^{\oplus}C_A(t) J'=-J''(u^{-1})^{\dagger \oplus}J'' J''\sigma_z^{\oplus}{\mathcal C}_{t}^{(A)}J' J'(u^\dagger)^\oplus J'\\\sim -J''\sigma_z^{\oplus}{\mathcal C}_{t}^{(A)}J',
\end{multline}
as we wanted to show, where the last step follows from the fact that $J''(u^{-1})^{\dagger \oplus}J''=(J'(u^\dagger)^\oplus J')^{-1}$. 
By using~\eqref{eq:can_to_lad},\eqref{eq:first},\eqref{eq:last}, we obtain that 
\begin{equation}
	i\Omega J \Gamma_A(t)J\sim i\Omega J \overline{\Gamma}^{(A)}_t J,\quad\mathrm{with}\,\, \overline{\Gamma}^{(A)}_t:=R^\oplus {\mathcal C}_{t}^{(A)} (R^\dagger)^\oplus
\end{equation}
From~\eqref{eq:r} one has that, as anticipated, $\overline{\Gamma}^{(A)}_t$ is 
the two-point correlators of the position-momentum operators $x^{(k)}$ and $p^{(k)}$ defined from the quasiparticles operators $\eta_j$ as 
\begin{equation}
\label{eq:x-def}
x^{({k)}}=\frac{\eta_j(k)+\eta_j^\dagger(k)}{\sqrt{2}},\quad p_j^{(k)}=-i\frac{\eta_j(k)-\eta_j(k)^\dagger}{\sqrt{2}}.
\end{equation}
Thus, the correlator  $\Gamma_{\mathcal{A}_1\cup\mathcal{A}_2}(x,k,t)$ of the 
generic $n$-quasiparticle multiplet  with quasimomentum $k$ and originated at $x$ is 
\begin{equation}
\label{eq:gammak}
\Gamma_{\mathcal{A}_1\cup\mathcal{A}_2}(x,k,t):=R^{\oplus |\mathcal{A}_1\cup\mathcal{A}_2|} \widetilde{{\mathcal C}}^{(\mathcal{A}_1\cup\mathcal{A}_2)}_{x,k,t}(t) (R^\dagger)^{\oplus |\mathcal{A}_1\cup\mathcal{A}_2|},
\end{equation}
and we have $\overline{\Gamma}_t^{(A)}=\bigoplus_{x,k}\Gamma_{\mathcal{A}_1\cup\mathcal{A}_2}(x,k,t)$. 
In terms of $x^{(k)}$ and $p^{(k)}$ the negativity of the multiplet resembles the standard definition of the negativity (see Ref.~\cite{Eisert_2018neg}). 
We stress again that the operators~\eqref{eq:x-def} satisfy canonical commutation relations and are \textit{not} Fourier transforms of any set of position-momentum operators, since, for example, they are self-adjoint, while Fourier transforms of position-momentum operators are not. 

There is an alternative way to justify Eqs.~\eqref{eq:trace-hydro} and~\eqref{eq:trace-hydro-diss}. One can notice that the identities of the form $f(\rho_A(t))=\mathrm{Tr}\left[\mathcal{F}\left(C_A(t)\right)\right]$ holding for Gaussian states usually rest only upon the assumption that the local operators satisfy canonical (anti)commutation relations. Thus, choosing another basis implies simply that $\mathcal{F}$ has to be applied to the corresponding two-point function, that in our case is exactly $\mathcal{C}_t^{(A)}$ . 

\section{Detailed calculations for our models}
\label{app:detail}

\subsection{Quench of the magnetic field in the Kitaev chain with gain and loss dissipation}
\label{app:kitaev}

First, let us focus on the key fermionic model we presented in the main text. Namely, let us consider a quench of the magnetic field in the Kitaev chain. The Hamiltonian reads as 
\begin{equation}
\label{eq:xy_app}
H_{K}=-\frac{1}{2}\sum_{j=1}^{L} \left(c_j^\dagger c_{j+1}+\delta c_j^\dagger c_{j+1}^\dagger + h. c. -2h c_j^\dagger c_j\right).
\end{equation}
We assume periodic boundary conditions, i.e., $c_{L+1}\equiv c_1$. At $t<0$, the system is in the ground state of $H_K$ with $h=h_0$. At $t=0$, the field is instantly quenched to $h=h_1$. Let us first consider the unitary case.
$H_K$ can be diagonalized by means of a Fourier transform 
$c_k=\frac{1}{\sqrt{L}}\sum_{j=1}^L e^{ikj} c_j$ followed by a Bogoliubov transformation, i.e.,
\begin{equation}
\label{eq:bogo_ferm_app}
\begin{pmatrix}
\eta_k\\\eta_{-k}^\dagger
\end{pmatrix}=\begin{pmatrix}
\cos \left(\Delta_{k,h}/2\right) &-i\sin\left(\Delta_{k,h}/2\right)\\
-i\sin\left(\Delta_{k,h}/2\right)&\cos \left(\Delta_{k,h}/2\right)
\end{pmatrix} \begin{pmatrix}
c_k\\c_{-k}^\dagger
\end{pmatrix},
\end{equation}
with the Bogoliubov angle $\Delta_{k,h}$ satisfying
\begin{align}
\label{eq:bogo_angle}
& \cos \Delta_{k,h}=\frac{h-\cos k}{\sqrt{(h-\cos k)^2+
		\delta^2\sin^2 k}} \\ 
& \sin \Delta_{k,h}=\frac{\delta \sin k}{\sqrt{(h-\cos k)^2+\delta^2\sin^2 k}}.
\end{align}
In the thermodynamic limit $L\to\infty$, we can rewrite:
\begin{equation}
\label{eq:xy_diag_app}
H_{K}=\int_{0}^\pi \frac{dk}{2\pi} \varepsilon (k)\left( \eta_k^\dagger \eta_k + \eta_{-k}^\dagger \eta_{-k} -1\right), 
\end{equation}
with $\varepsilon(k)$ given as 
\begin{equation}
	\label{eq:kit-en}
	\varepsilon(k):=\sqrt{(h-\cos k)^2+\delta^2\sin^2 k}
\end{equation}
The initial state is the ground state of $H_{K}$ with the 
field $h=h_0$. Let us call $\eta_k^{(0)}$ the eigenmodes 
for $h=h_0$. The ground state then reads $|\Omega\rangle\propto 
\prod_{k=-\pi}^\pi \eta_k^{(0)} |0\rangle$. From 
Eq.~\eqref{eq:bogo_ferm_app}, we see that the post-quench 
eigenmodes $\eta_k$ satisfy:
\begin{equation}
\label{eq:xy_chmodes}
\begin{pmatrix}
\eta_k\\\eta_{-k}^\dagger
\end{pmatrix}=
M
\begin{pmatrix}
\eta_k^{(0)}\\\left.\eta_{-k}^{(0)}\right.^\dagger
\end{pmatrix},
\end{equation}
where we defined $M$ as 
\begin{equation}
M=\begin{pmatrix}
\cos \left(\theta_{h_1,h_0}(k)/2\right) &-i\sin\left(\theta_{h_1,h_0}(k)/2\right)\\
-i\sin\left(\theta_{h_1,h_0}(k)/2\right)&\cos \left(\theta_{h_1,h_0}(k)/2\right)
\end{pmatrix} \end{equation}
with $\theta_{h',h}(k):=\Delta_{k,h'}-\Delta_{k,h}$. In the 
following, we drop the subscripts $h_0$, $h_1$ for the sake of simplicity, and write $\theta_k$ for $\theta_{h_1,h_0}(k)$. 
As explained above, we identify the pairs of entangled excitations, which are responsible of entanglement spreading, 
as $\eta_1^\dagger(k)=\eta_k^\dagger$, $\eta_2^\dagger(k)=\eta_{-k}^\dagger$, $k\in[0,\pi]$, 
with the dispersions $\epsilon_1(k)=\epsilon_2(k)=\varepsilon(k)$ that we obtain 
from~\eqref{eq:xy_diag_app}. The group velocities of the quasiparticles are then
$v_1(k)=-v_2(k)=\varepsilon'(k)$ (cf.~\eqref{eq:kit-en}). 

By using~\eqref{eq:xy_diag_app} and~\eqref{eq:xy_chmodes}, 
one obtains that the two-point function is block-diagonal 
in the quasimomentum space, with the block encoding the correlation 
between the species (Eq.~\eqref{eq:corr} in the main text) reading
\begin{widetext}
\begin{equation}
\label{eq:xy_un_blockdiag}
\mathcal{C}_{x,k}(t)=\frac{1}{2}\begin{pmatrix}
1-\cos \theta_k&0&0&i\sin\theta_k e^{2i\varepsilon(k)t}\\
0&1+\cos\theta_k&i\sin\theta_ke^{-2i\varepsilon(k)t}&0\\
0&-i\sin\theta_ke^{2i\varepsilon(k)t}&1-\cos\theta_k&0\\
-i\sin\theta_ke^{-2i\varepsilon(k)t}&0&0&1+\cos\theta_k
\end{pmatrix}.
\end{equation}
\end{widetext}
Notice that in~\eqref{eq:xy_un_blockdiag} we keep the phases $e^{2i\varepsilon(k)t}$ associated 
with the unitary evolution, thus it represents the time evolution of $\mathcal{C}_{x,k}$ in Eq.~\eqref{eq:corr} of the main text. 
	We then add gain and loss dissipation, defined by the two families of Lindblad operators $L_{+,j}=\sqrt{\gamma_+}c_j^\dagger$, $L_{-,j}=\sqrt{\gamma_-}c_j$ in Eq.~\eqref{eq:lindblad_state}, where $\gamma_\pm$ are the gain and loss rates. We straightforwardly obtain that the Fourier transforms of $A$ and $B$ (cf.~\eqref{eq:lindblad_lin}) read
	\begin{equation}
	\label{eq:gainloss_ferm}
	\hat{A}_k=\frac{\gamma_++\gamma_-}{4} \mathds{1}_2, \qquad \qquad \hat{B}_k=i\frac{\gamma_+-\gamma_-}{4}\sigma_y.
	\end{equation}
In the presence of a homogeneous dissipation, the two-point correlation function 
remains block-diagonal with respect of the two families $\eta_1^\dagger (k), 
\eta_2^\dagger (k)$. Indeed, a homogeneous dissipation can at most mix the modes $\eta_1,\eta_2^\dagger$. 
However, this does not happen for gain and loss dissipation, for which the modes that propagate 
entanglement remain the same as the eigenmodes of the Hamiltonian. This is due to the fact that 
$\hat{A}_k\propto \mathds{1}_{2}$ (cf.~\eqref{eq:chi_trinv}). To derive the time-dependent 
correlator $\widetilde{\mathcal{C}}_{x,k,t}(t)$, it is easier to first go to the Majorana basis, by defining the Majorana 
operators $r_j$ as $r_{2j-1}=c_j+c_j^\dagger$, $r_{2j}=i(c_j-c_j^\dagger)$. 
The correlation matrix $\Gamma_{m,n}:=\langle[r_m,r_n]/2\rangle$, which contains the same information as 
$\widetilde{\mathcal{C}}_{x,k,t}(t)$, satisfies the evolution equation~\cite{Carollo2022entdiss}
\begin{equation}
\label{eq:corrt_real_ferm}
\Gamma(t)=U(t)\Gamma(0) U^T(t) + 4i \int_{0}^{t} du U(t-u) B U^T(t-u),
\end{equation}
where $U(t)=e^{-(4ih+2A)t}$, with $A$, $B$ and $h$ defined as in~\eqref{eq:lindblad_op}. 
Translational invariance allows us to rewrite Eq.~\eqref{eq:corrt_real_ferm} 
into a time-evolution equation for the Fourier transform $\hat{g}_k$ of the correlation 
matrix as 
\begin{equation}
\label{eq:corrt_k_ferm_app}
\hat{g}_k=\hat{u}_k
\hat{g}_k(0) \hat{u}_{-k}^T 
+ 4i \int_{0}^{t} du \; \hat{u}_k(t-u) \hat{B}_k \hat{u}_{-k}^T(t-u),
\end{equation}
where $\hat{u}_k(t)=e^{-(4i\hat{h}_k+2\hat{A}_k)t}$, with $\hat{h}_k$, $\hat{A}_k$ 
and $\hat{B}_k$ the Fourier transforms  of $h$, $A$, $B$, respectively. 
The Fourier transform $\hat{g}_k$ of the correlation matrix reads
\begin{equation}
\label{eq:symbol_majorana_app}
\hat{g}_{k}=\begin{pmatrix}
\langle r_{o,k} r_{o,-k} \rangle -1&\langle r_{o,k} r_{e,-k} \rangle\\\langle r_{e,k} r_{o,-k} \rangle&\langle r_{e,k} r_{e,-k} \rangle-1
\end{pmatrix}.
\end{equation}
Here $r_{o,k}:=1/\sqrt{L}\sum_{j=1}^L e^{ikj} r_{2j-1}=c_k+c_{-k}^\dagger$ and 
$r_{e,k}:=1/{\sqrt{L}}\sum_{j=1}^L e^{ikj} r_{2j}=i(c_k-c_{-k}^\dagger)$ the 
Fourier transforms of odd and even Majorana fermions $r_{2j-1}$ and $r_{2j}$, respectively. We obtain 
$\widetilde{\mathcal{C}}_{x,k}(t)$ by solving Eq.~\eqref{eq:corrt_k_ferm_app} in the thermodynamic limit $L\to \infty$, and then changing basis as
\begin{widetext}
\begin{equation}
\label{eq:corrt_k_chbasis}
\widetilde{{\mathcal C}}_{x,k}(t)=\frac{1}{4}\begin{pmatrix}
0&0&m_k&i m_k^*\\
m_k^*&m_k^*&0&0\\
-im_k&im_k&0&0\\
0&0&m_k&-im_k^*
\end{pmatrix}\begin{pmatrix}
\hat{g}_k(t)+\mathds{1}_{2\times 2}&\mathbf{0}_{2\times 2}\\\mathbf{0}_{2\times 2}&\hat{g}_{-k}(t)+\mathds{1}_{2\times 2}
\end{pmatrix} \begin{pmatrix}
0&m_k&im_k^*&0\\
0&m_k&-im_k^*&0\\
m_k^*&0&0&m_k^*\\
-im_k&0&0&im_k
\end{pmatrix}, \quad m_k:=e^{\frac{i}{2}\Delta_{k,h_1}}.
\end{equation}
\end{widetext}
In~\eqref{eq:corrt_k_chbasis}, translation invariance ensures that $\langle r_{o,k} r_{e,k}\rangle=0$, 
and $\Delta_{k,h_1}$ is the Bogoliubov angle defined in~\eqref{eq:bogo_angle}. 
The time evolution of $\hat{g}_k(t)$ was computed in~\cite{Alba2022entdiss} for 
generic homogeneous linear dissipation in the weakly-dissipative hydrodynamic limit 
$t\to\infty$, $\gamma\to0$ with constant $\gamma t$, where $\gamma$ is 
the strength of the dissipation. In particular, for gain and loss dissipation one has 
\begin{multline}
\label{eq:corrt_k_xy}
\hat{g}_k(t)=\left(\cos\theta_k\sigma_y^{(k)}-\sin\theta_k\sigma_x^{(k)} 
e^{2i\varepsilon(k)t\sigma_y}\right)\lambda_t\\
+\left(1-\lambda_t\right)\left(1-2n_\infty\right)\cos \Delta_{k,h_1} \sigma_y^{(k)},
\end{multline}
where $\sigma_a,\; a=x,y,z$ are the Pauli matrices, and we define $\sigma_a^{(k)}:=
e^{i\Delta_{k,h_1}\sigma_z/2}\sigma_a e^{-i\Delta_{k,h_1}\sigma_z/2}$, 
$\lambda_t:=e^{-(\gamma_++\gamma_-)t}$, $n_\infty:=\gamma_+/(\gamma_++\gamma_-)$. 
Plugging~\eqref{eq:corrt_k_xy} into~\eqref{eq:corrt_k_chbasis}, we obtain 
\begin{widetext}
\begin{equation}
\label{eq:corrt_xy_app}
\widetilde{\mathcal{C}}_{x,k}(t)=\frac{1}{2}\begin{pmatrix}
1-A_{k,t}&0&0&-iB_{k,t}e^{2i\varepsilon(k)t}\\
0&1+A_{k,t}&-iB_{k,t}e^{-2i\varepsilon(k)t}&0\\
0&iB_{k,t}e^{2i\varepsilon(k)t}&1-A_{k,t}&0\\
iB_{k,t}e^{-2i\varepsilon(k)t}&0&0&1-A_{k,t}
\end{pmatrix},
\end{equation}
\end{widetext}
where
\begin{multline}
\label{eq:AkBk}
A_{k,t}:=\lambda_t \cos (\Delta_{k,h_1}-\Delta_{k,h_0})\\+(1-\lambda_t)(1-2n_\infty)\cos \Delta_{k,h_1}
\end{multline}
and 
\begin{equation}
B_{k,t}:=\lambda_t \sin (\Delta_{k,h_1}-\Delta_{k,h_0}),
\end{equation}
which is Eq.~\eqref{eq:corrt_xy} in the main text. It is now straightforward to determine 
the negativity between the two quasiparticles forming an entangled pair.

\subsection{Quench from the N\'eel state in the tight-binding chain with gain and loss dissipation}
\label{app:tb-neel}

As a second example, we consider a quench in the tight-binding chain with the gain and loss dissipation~\eqref{eq:gainloss_ferm}. The initial state we choose is the fermionic N\'eel state
\begin{equation}
\label{eq:neel}
	|N\rangle=\prod_{j=1}^{L/2}c_{2j-1}^\dagger |0\rangle.
\end{equation}
This setting is particularly interesting because the exact expression for the negativity in the
weakly-dissipative hydrodynamic limit is already known, as it was calculated in~\cite{alba2022logarithmic}. We now
show that our quasiparticle framework reproduces correctly and in a simpler way this result,
whose \textit{ab initio} analytical derivation is long and involved.
The tight-binding chain is obtained as the limiting case $h=\delta=0$ of the Kitaev 
chain~\eqref{eq:xy_app}. The tight-binding  Hamiltonian thus reads
\begin{equation}
\label{eq:xx}
H=-\frac{1}{2}\sum_{j=1}^{L} \left(c_j^\dagger c_{j+1}+ c_{j+1}^\dagger c_j\right),
\end{equation}
and we assume periodic boundary conditions.
Now, the initial state has a weaker translational
symmetry than the Kitaev chain, as it presents only a two-site translation invariance. In principle,
as suggested by Eq.~\eqref{eq:corrt_k_ferm_app}, one should thus expect a $4\times4$ correlation matrix 
$\hat{g}_k$ encoding the correlations between the Fourier modes 
$c_{k}^\dagger$, $c_{-k}^\dagger, c_{k-\pi}^\dagger$, $c_{-k+\pi}^\dagger$, with $k\in \left[0,\pi/2\right]$.
However, the simple structure of the tight-binding chain and of the N\'eel state ensures that 
some of these correlations are zero. This allows us to describe the dynamics of the negativity 
in terms of pairs. Indeed, the tight-binding chain can be diagonalized in terms of the Fourier modes, in the thermodynamic limit $L\to\infty$, as 
\begin{equation}
\label{eq:xx_diag}
H=\int_0^\pi \varepsilon(k) \left(c_k^\dagger c_k-c_{k-\pi}^\dagger c_{k-\pi}\right),
\end{equation}
with $\varepsilon(k)=-\cos k$. In~\eqref{eq:xx_diag} $c_k^\dagger,c_{k-\pi}^\dagger$ are the two species of entangled quasiparticles.
Precisely, we define $\eta_1^\dagger(k)=c_k^\dagger$, $\eta_2^\dagger(k)=c_{k-\pi}^\dagger$. The modes 
propagate with opposite velocities $v_1(k)=-v_2(k)=\varepsilon'(k)=\sin k$. Notice that 
$v_j(k)=\varepsilon_j'(k)$, with $\varepsilon_j(k)$ the dispersions of the eigenmodes,
unlike the case of the Kitaev chain, in which $v_2(k)=-\varepsilon_2'(k)$, due 
to the different folding of the Brillouin zone. The correlation function in  quasimomentum space is block-diagonal and reads
\begin{equation}
\label{eq:corrt_neel_unitary}
\mathcal{C}_{x,k}(t)=\frac{1}{2}\begin{pmatrix}
1&0&-e^{2i\varepsilon(k)t}&0\\
0&1&0&e^{-2i\varepsilon(k)t}\\
-e^{-2i\varepsilon(k)t}&0&1&0\\
0&e^{2i\varepsilon(k)t}&0&1\\
\end{pmatrix}. 
\end{equation}
In the presence of gain and loss dissipation, the block structure of~\eqref{eq:corrt_neel_unitary} remains 
the same. Indeed, for gain and loss dissipation the Lindblad equation~\eqref{eq:lindblad_op} can be rewritten, for $L\to \infty$, as
\begin{multline}
\label{eq:lindblad_gainloss}
i\hat{\mathcal{L}}^\dagger[O]=i[H,O]+ \int_{-\pi}^\pi \frac{dk}{2\pi}\Big[ \gamma_+\left(c_k O c_k^\dagger-\frac{1}{2}\left\{c_k c_k^\dagger, O \right\}\right)\\+\gamma_- \left(c_k^\dagger O c_k-\frac{1}{2} \left\{c_k^\dagger c_k,O\right\}\right)\Big].
\end{multline}
It is easy to derive the time evolution of the two-point correlators as 
\begin{equation}
\label{eq:cct_gainloss}
\begin{gathered}
\frac{d}{dt} \langle c_k^\dagger c_h \rangle=\left[i \left(\varepsilon(k)-\varepsilon(h)\right)-\left(\gamma_++\gamma_-\right)\right] \langle c_k^\dagger c_h \rangle + \gamma_+ \delta_{k,h},\\
\frac{d}{dt} \langle c_k^\dagger c_h^\dagger \rangle=\left[i \left(\varepsilon(k)+\varepsilon(h)\right)-\left(\gamma_++\gamma_-\right)\right] \langle c_k^\dagger c_h^\dagger \rangle.
\end{gathered}
\end{equation} 
From~\eqref{eq:cct_gainloss}, it is apparent that the two bilinears $c_k^\dagger c_k$ and 
$c_{k-\pi}^\dagger c_{k-\pi}$ are eigenfunctions of the Liouvillian in the sense 
of~\eqref{eq:eigen_diss_ferm1}. This also means that the propagating modes remain the same as in the unitary case. 
In contrast with the case of systems with one-site translation invariance (see Appendix~\ref{app:kitaev}), 
the eigenmodes are not of the form $c^\dagger_k,c^\dagger_{-k}$. 
However, comparing with the unitary case, it is apparent that the other bilinear we have to consider 
to extract the dispersion is $[c_k^\dagger, c_{k-\pi}]$. The velocities of the modes $\eta_1,\eta_2$ are readily obtained from the dynamics of $[c_k^\dagger,c_{k-\pi}]$.
Alternatively, considering $\left[c_k, c_{-k}\right]$ and $\left[c_{k-\pi}, c_{\pi-k}\right]$ would work as well, since, as suggested by the symmetry, this quench actually produces \textit{quadruplets} that are split into two superposed and uncorrelated pairs each.
One obtains that the velocities are the same as in the unitary case, i.e., 
$v_1(k)=-v_2(k)=\varepsilon'(k)$. 
The time-evolved correlation matrix reads
\begin{widetext}
\begin{equation}
\label{eq:corrt_neel_diss}
\widetilde{\mathcal{C}}_{x,k}(t):=\langle \eta^\dagger_i\eta_j\rangle= 
\mathcal{C}_{x,k}(t) e^{-(\gamma_++\gamma_-)t}+
\left(1-e^{-(\gamma_++\gamma_-)t}\right)\begin{pmatrix}
n_\infty&0&0&0\\
0&1-n_\infty&0&0\\
0&0&n_\infty&0\\
0&0&0&1-n_\infty
\end{pmatrix},
\end{equation}
\end{widetext}
where $\mathcal{C}_{x,k}(t)$ is the correlation matrix in the case without dissipation 
(cf.~Eq.~\eqref{eq:corrt_neel_unitary}), and $n_\infty:=\gamma_+/({\gamma_++\gamma_-})$ as above. 
After removing the phases $e^{\pm2i\varepsilon(k)t}$ from $\widetilde{\mathcal{C}}_{x,k}(t)$ in~\eqref{eq:corrt_neel_diss} and passing to $\widetilde{\mathcal{C}}_{x,k,t}(t)$, we use Eq.~$(11)$ of the main text, and we obtain the negativity of a pair of shared quasiparticles as 
\begin{multline}
\label{eq:neel_neg_app}
e_2(k,t)=\ln\Big( \lambda_t+2g_\times(1-\lambda_t^2) \\+\sqrt{(1-\lambda_t)^2[
	(1-\lambda_t)g_+ +\lambda_t]^2+\lambda_t^2}\Big),
\end{multline}
where we define $g_\times:=(\gamma^+\gamma^-)/((\gamma^++\gamma^-)^2)$, 
$g_+:=((\gamma^+)^2+(\gamma^-)^2)/((\gamma^++\gamma^-)^2)$ and $\lambda_t:=e^{-(\gamma_++\gamma_-)t}$ as above. It is straightforward to 
check that Eq.~\eqref{eq:neel_neg_app} coincides with the result of the \textit{ab initio} derivation 
in~\cite{alba2022logarithmic}. Notice that the analytical derivation required very long and involved calculations, therefore our method represents a remarkable simplification.

\subsection{Quench from the dimer state in the tight-binding chain with gain and loss dissipation}
\label{app:dimer}

Let us consider a different initial state in the tight binding chain~\eqref{eq:xx} with the gain and loss
dissipation~\eqref{eq:gainloss_ferm}, specifically, the dimer state
\begin{equation}
\label{eq:dimer}
	|D\rangle=\prod_{j=1}^{L/2} \frac{1}{\sqrt{2}}\left(c_{2j-1}^\dagger +c_{2j}^\dagger  \right)|0\rangle
\end{equation}
It is easy to verify that the eigenmodes are the same as for the N\'eel state, and in the unitary case the correlation matrix is given as 
\begin{widetext}
\begin{equation}
\label{eq:corrt_dimer_unitary}
\mathcal{C}_{x,k}(t)=\frac{1}{2}\begin{pmatrix}
1+\cos k&0&i\sin ke^{2i\varepsilon(k)t}&0\\
0&1-\cos k&0&i\sin k e^{-2i\varepsilon(k)t}\\
-i\sin k e^{-2i\varepsilon(k)t}&0&1-\cos k&0\\
0&-i \sin k e^{2i\varepsilon(k)t}&0&1+\cos k\\
\end{pmatrix}.
\end{equation}
\end{widetext}
Similarly to the N\'eel case, the velocities are the same as in the 
unitary case and are left unmodified by the gain and loss dissipation~\eqref{eq:gainloss_ferm}. In the dissipative case, we obtain 
\begin{multline}
\label{eq:corrt_dimer_diss}
\widetilde{\mathcal{C}}_{x,k}(t)=\mathcal{C}_{x,k}(t) 
e^{-(\gamma_++\gamma_-)t}\\
+\left(1-e^{-(\gamma_++\gamma_-)t}\right)\begin{pmatrix}
n_\infty&0&0&0\\
0&1-n_\infty&0&0\\
0&0&n_\infty&0\\
0&0&0&1-n_\infty
\end{pmatrix},
\end{multline}
where $\mathcal{C}_{x,k}(t)$ is the correlator in the unitary case, Eq.~\eqref{eq:corrt_dimer_unitary}.
We remove the phases associated with the unitary part of the dynamics and pass to $\widetilde{\mathcal{C}}_{x,k,t}(t)$, and then we 
apply Eq.~\eqref{eq:trace-hydro-diss} of the main text in the case of entangled pairs.
We finally obtain
\begin{multline}
\label{eq:dimer_neg_app}
e_2(k,t)=\ln\Big (\lambda_t+2g_\times(1-\lambda_t^2) \\+\sqrt{(1-\lambda_t)^2[
	(1-\lambda_t)g_+ +\lambda_t]^2+\lambda_t^2\sin^2 k}\Big),
\end{multline}
where $g_\times$ and $g_+$ are the same as in~\eqref{eq:neel_neg_app}. Eq.~\eqref{eq:dimer_neg_app} is the 
same formula as discussed in the main text.

\subsection{Quench of the mass in the harmonic chain with gain and loss dissipation}
\label{app:hc}

As a case-study of bosonic systems, we consider a mass quench in  the harmonic chain
\begin{equation}
\label{eq:harmonic_app}
H_{h}=\frac{1}{2}\sum_{j=1}^{L}\left(\frac{p_j^2}{m}+m w^2 x_j^2 + \mathcal{K}(x_{j+1}-x_j)^2 \right)
\end{equation}
with periodic boundary conditions. The chain is prepared in the ground state of~\eqref{eq:harmonic_app} for $m=m_0$. 
At $t=0$, the mass is suddenly changed to the value $m=m_1$. The local canonical 
operators $x_j,p_j$ play the same role as the Majorana operators for fermionic systems 
in~\eqref{eq:lindblad_lin}, by identifying $r_{2j-1}=x_j$, $r_{2j}=p_j$. 
It is useful to summarize their canonical commutation relations as:
\begin{equation}
\label{eq:comm}
\left[r_m,r_n\right]=i \Omega_{m,n}, \quad \text{with} \quad \lfloor \Omega_{i,j}\rceil =\begin{pmatrix}
0&1\\-1&0
\end{pmatrix} \delta_{i,j}=i\sigma_y \delta_{i,j}.
\end{equation}
Notice that in the bosonic case the commutation relations are preserved by the simplectic transformations $S$, i.e., real linear transformations such that $S\Omega S^T=\Omega$. The canonical operators can be combined to form the local ladder operators $b_j=(x_j+i p_j)/\sqrt{2}$, $b_j^\dagger=(x_j-i p_j)/\sqrt{2}$, satisfying $[b_i,b_j^\dagger]=\delta_{i,j}$. For a Gaussian state, all the information is contained in the two-point function $\Gamma_{m,n}:=\langle \left\{r_m, r_n\right\} \rangle$. From Eqs.~\eqref{eq:lindblad_op} and~\eqref{eq:lindblad_lin}, it follows that $\Gamma$ satisfies the time-evolution equation (see~\cite{Carollo2022entdiss})
\begin{equation}
\label{eq:corrt_real_bos}
\Gamma(t)=W(t)\Gamma(0) W(t)^T + 2 \int_{0}^{t} du W(t-u) \Omega A \Omega^T W(t-u)^T.
\end{equation}
Here $W(t)=e^{\Omega(2h+B)t}$, and $A$, $B$, $h$ are defined as in~\eqref{eq:lindblad_lin}. By exploiting the translational invariance, we
can rewrite Eq.~\eqref{eq:corrt_real_bos} as an evolution equation for the Fourier transform  
$\hat{g}_k$ of the two-point function, reading 
\begin{equation}
\label{eq:symbol_bos}
\hat{g}_{k}=2\begin{pmatrix}
\langle x_k x_{-k} \rangle &\langle x_k p_{-k} \rangle\\\langle p_k x_{-k} \rangle&\langle p_k p_{-k} \rangle
\end{pmatrix} + \sigma_y.
\end{equation}
Here $x_k$ and $p_k$ are the Fourier transforms of the canonical operators, defined as $x_k:=1/{\sqrt{L}} \sum_j e^{ikj} x_j=(b_k+b_{-k}^\dagger)/\sqrt{2}$ and 
$p_k:=1/{\sqrt{L}} \sum_j e^{ikj} p_j=-i(b_k-b_{-k}^\dagger)/\sqrt{2}$, where 
$b_k,b_k^\dagger$ are the Fourier modes associated to the ladder operators.
In terms of $\hat{g}_{k}$, Eq.~\eqref{eq:corrt_real_bos} becomes~\cite{Carollo2022entdiss}
\begin{equation}
\label{eq:corrt_k_bos}
\hat{g}_{k}(t)=\hat{W}_k(t)\hat{g}_k(0) 
\hat{W}^\dagger_{k}(t) + 2 \int_{0}^{t} du \hat{W}_k(t-u) \sigma_y \hat{A}_k \sigma_y 
\hat{W}_{k}^\dagger(t-u),
\end{equation}
where $\hat{W}_k(t)=e^{i\sigma_y(2\hat{h}_k+\hat{B}_k)t}$.
In the thermodynamic limit $L\to\infty$, the Fourier transform of the Hamiltonian reads
\begin{equation}
\label{eq:harm_hk}
\hat{h}_k=\begin{pmatrix}
2a(k)&0\\0&2b
\end{pmatrix},
\end{equation}
where we defined $a(k):=m_1w^2/4+\mathcal{K}\sin^2(k/2)$ and $b:=1/(4m_1)$.
The harmonic chain~\eqref{eq:harmonic_app} can be diagonalized, in the thermodynamic limit $L\to \infty$, as
\begin{multline}
\label{eq:harm_diag}
H_h=\int_{-\pi}^\pi \frac{dk}{2\pi} \omega(k) \left(\eta_k^\dagger \eta_k +\frac{1}{2}\right)\\
=\int_0^\pi \frac{dk}{2\pi} \omega(k) \left(\eta_k^\dagger \eta_k +\eta_{-k}^\dagger \eta_{-k}+1\right), 
\end{multline}
with 
\begin{equation}
\omega(k):=\sqrt{w^2+\frac{4\mathcal{K}}{m}\sin^2(k/2)}.
\end{equation}
This can be done by means of a Fourier transform followed by the transformation
\begin{equation}
\label{eq:bogo_bos}
\begin{pmatrix}
\eta_{k}\\\eta_{-k}^\dagger
\end{pmatrix}=\frac{1}{2}\begin{pmatrix}
z_m(k)+\frac{1}{z_m(k)}&z_m(k)-\frac{1}{z_m(k)}\\
z_m(k)-\frac{1}{z_m(k)}&z_m(k)+\frac{1}{z_m(k)}
\end{pmatrix} \begin{pmatrix}
b_k\\b_{-k}^\dagger
\end{pmatrix},
\end{equation}
where 
\begin{equation}
	z_m(k):=\sqrt[4]{m^2w^2 + 4 m \mathcal{K} \sin^2(k/2)}, 
\end{equation}
that can be seen as a bosonic version of the Bogoliubov transformation~\eqref{eq:bogo_ferm_app}. The modes $\eta_k$ satisfy the same commutation relations as the $b_k$. Indeed, the latter can be summarized as
\begin{equation}
\label{eq:ladder_comm}
\left[\begin{pmatrix}
b_k\\b_{-k}^\dagger
\end{pmatrix},\begin{pmatrix}
b_{-h}&b_h^\dagger
\end{pmatrix}\right]=\begin{pmatrix}
0&1\\-1&0
\end{pmatrix}\delta_{k,h}=i\sigma_y \delta_{k,h},
\end{equation}
thus they are preserved by the transformation~\eqref{eq:bogo_bos}, which is symplectic and even in $k$.
Our initial state is the ground state $|\Omega\rangle$ of $H_h$ with $m=m_0$. 
Let us call $\eta_k^{(0)}$ the diagonal modes of $H_h$ for $m=m_0$. They annihilate the initial state. Eq.~\eqref{eq:bogo_bos} ensures that the post-quench modes $\eta_k$ satisfy
\begin{equation}
\label{eq:harm_chmodes}
\begin{pmatrix}
\eta_{k}\\\eta_{-k}^\dagger
\end{pmatrix}=\frac{1}{2}\begin{pmatrix}
\frac{z(k)^2+z_0(k)^2}{z(k)z_0(k)}&\frac{z(k)^2-z_0(k)^2}{z(k)z_0(k)}\\
\frac{z(k)^2-z_0(k)^2}{z(k)z_0(k)}&\frac{z(k)^2+z_0(k)^2}{z(k)z_0(k)}
\end{pmatrix} \begin{pmatrix}
\eta_k^{(0)}\\\left.\eta_{-k}^{(0)}\right.^\dagger
\end{pmatrix},
\end{equation}
where we call $z_0(k):=z_{m_0}(k)$ and $z(k):=z_{m_1}(k)$ (see~\eqref{eq:bogo_bos}). Now, we can identify 
the two families of entangled quasiparticles as $\eta_1^\dagger(k)=\eta_k^\dagger$, 
$\eta_2^\dagger(k)=\eta_{-k}^\dagger$. In terms of these quasiparticles the two-point function is 
block-diagonal in the quasimomentum space, as for the fermionic models. The associated velocities 
are $v_1(k)=-v_2(k)=\omega'(k)$. In the unitary case, we easily find
\begin{widetext}
\begin{equation}
\label{eq:harm_corrt_unitary}
\mathcal{C}_{x,k}(t)=\frac{1}{4}\begin{pmatrix}
\frac{(z(k)^2-z_0(k)^2)^2}{z(k)^2z_0(k)^2}&0&0&\frac{z(k)^4-z_0(k)^4}{z(k)^2z_0(k)^2}e^{2i\omega(k)t}\\
0&\frac{(z(k)^2+z_0(k)^2)^2}{z(k)^2z_0(k)^2}&\frac{z(k)^4-z_0(k)^4}{z(k)^2z_0(k)^2}e^{-2i\omega(k)t}&0\\
0&\frac{z(k)^4-z_0(k)^4}{z(k)^2z_0(k)^2}e^{2i\omega(k)t}&\frac{(z(k)^2-z_0(k)^2)^2}{z(k)^2z_0(k)^2}&0\\
\frac{z(k)^4-z_0(k)^4}{z(k)^2z_0(k)^2}e^{-2i\omega(k)t}&0&0&\frac{(z(k)^2+z_0(k)^2)^2}{z(k)^2z_0(k)^2}
\end{pmatrix}.
\end{equation}
\end{widetext}
We then add a gain and loss dissipation. It is defined as in the fermionic case by two families of Lindblad operators $L_{+,j}=\sqrt{\gamma_+}b_j^\dagger$, $L_{-,j}=\sqrt{\gamma_-}b_j$ in Eq.~\eqref{eq:lindblad_state}, where $\gamma_\pm$ are the gain and loss rates. The different definition of the $r_j$ in the bosonic case yields Fourier transforms of $A$ and $B$ (cf.~\eqref{eq:lindblad_lin}) that are slightly different from the fermionic case, reading
	\begin{equation}
	\label{eq:gainloss_bos}
	\hat{A}_k=\frac{\gamma_++\gamma_-}{2} \mathds{1}_2, \qquad \qquad \hat{B}_k=i\frac{\gamma_--\gamma_+}{2}\sigma_y.
	\end{equation}
As in the fermionic case, the block structure in~\eqref{eq:harm_corrt_unitary} is preserved when a 
linear, homogeneous dissipation is added. Indeed, the translation invariance ensures 
that the only non-zero correlations are those contained in~\eqref{eq:symbol_bos}.
Thus, dissipation can at most mix the propagating modes $\eta_1(k), \eta_2^\dagger(k)$. 
This, however, does not happen for the gain and loss dissipation, which preserves the eigenmodes of the 
Hamiltonian as propagating modes, since $\hat{B}_k\propto \sigma_y$, and the propagation velocities (see the 
discussion in Appendix~\ref{app:eigen} above). One can obtain the time-dependent correlator $\widetilde{\mathcal{C}}_{x,k}(t)$ 
by solving the time-evolution equation~\eqref{eq:corrt_k_bos} for the Fourier 
transform~\eqref{eq:symbol_bos} and then changing basis as 
\begin{equation}
\label{eq:chbasis_harm}
\widetilde{\mathcal{C}}_{x,k}(t)=\frac{1}{2}M(k)
\begin{pmatrix}
\hat{g}_k(t)-\sigma_y&0_{2\times 2}\\
0_{2\times 2}&\hat{g}_{-k}(t)-\sigma_y
\end{pmatrix} M(k)^\dagger,
\end{equation}
with 
\begin{equation}
M(k):=\frac{1}{\sqrt{2}}\begin{pmatrix}
0&0&z(k)&-\frac{i}{z(k)}\\
z(k)&\frac{i}{z(k)}&0&0\\
z(k)&-\frac{i}{z(k)}&0&0\\
0&0&z(k)&\frac{i}{z(k)}
\end{pmatrix}. 
\end{equation}
In Eq.~\eqref{eq:chbasis_harm} the translation invariance ensures that $\langle x_k x_k \rangle=\langle p_k p_k \rangle=\langle x_k p_k \rangle=0$, and $z(k)$ is defined in~\eqref{eq:harm_chmodes}.
At $t=0$ the Fourier transform of the correlator 
is given as
\begin{multline}
\label{eq:harm_symbol_t0}
\frac{1}{2} \begin{pmatrix}
\hat{g}_k(0)-\sigma_y&0_{2\times 2}\\
0_{2\times 2}&\hat{g}_{-k}(0)-\sigma_y
\end{pmatrix}\\
=M^{-1}_0(k)\frac{1}{2}\begin{pmatrix}
\mathds{1}_{2\times 2}-\sigma_z &0_{2\times 2}\\
0_{2\times 2} & \mathds{1}_{2\times 2}-\sigma_z
\end{pmatrix} \left(M^\dagger_0(k)\right)^{-1},
\end{multline}
where $M_0(k)$ is the same matrix as $M(k)$ with $z_0(k)$ instead of $z(k)$, 
and we have used that the $\eta^{(0)}_j$ annihilate the initial state. 
By using Eq.~\eqref{eq:corrt_k_bos} with the initial condition~\eqref{eq:harm_symbol_t0}, we obtain 
\begin{widetext}
\begin{equation}
\label{eq:symbol_t_harm}
\begin{gathered}
\hat{g}_k(t)=\begin{pmatrix}
\left(\frac{\delta_t}{z_0^2}+\frac{1-\delta_t}{g}\right)\cos^2(\omega t) + \frac{1}{z^4}\left(z_0^2\delta_t +\frac{1-\delta_t}{g}\right)\sin^2(\omega t)&\frac{2}{\omega}\left(\frac{b-a}{g}+\delta_t\left(b z_0^2+\frac{a-b}{g}-\frac{a}{z_0^2}\right)\right)\sin(2\omega t)\\
\frac{2}{\omega}\left(\frac{b-a}{g}+\delta_t\left(b z_0^2+\frac{a-b}{g}-\frac{a}{z_0^2}\right)\right)\sin(2\omega t)&\left(z_0^2\delta_t +\frac{1-\delta_t}{g}\right)\cos^2(\omega t)+z^4	\left(\frac{\delta_t}{z_0^2}+\frac{1-\delta_t}{g}\right)\sin^2(\omega t)\\
\end{pmatrix} \delta_t +\\
+\begin{pmatrix}
-\frac{z^4+1}{2g z^4} +\frac{2\omega^2(z^4+1)/z^4+ g_-^2}{g(4\omega^2+g_-^2)\delta_t} +\frac{(z^4-1)(2\omega \sin(2\omega t)-g_-\cos(2\omega t))}{2z^4(4\omega^2+g_-^2)}&\frac{(z^4-1)(2\omega \cos(2\omega t)+g_-\sin(2\omega t)-2\omega\delta_{-t})}{2z^2(4\omega^2+g_-^2)}\\
\frac{(z^4-1)(2\omega \cos(2\omega t)+g_-\sin(2\omega t)-2\omega\delta_{-t})}{2z^2(4\omega^2+g_-^2)}&-\frac{z^4+1}{2g} +\frac{2\omega^2(z^4+1)+ g_-^2}{g(4\omega^2+g_-^2)\delta_t} +\frac{(z^4-1)(g_-\cos(2\omega t)-2\omega \sin(2\omega t))}{2(4\omega^2+g_-^2)}\\
\end{pmatrix} \delta_t.
\end{gathered}	
\end{equation}
\end{widetext}
Here we define $g_-:=\gamma_--\gamma_+$, $g:=(\gamma_--\gamma_+)/(\gamma_-+\gamma_+)$, 
$\delta_t:=e^{-g_-t}$, and $\omega(k)=4\sqrt{a(k)b}$ is defined in~\eqref{eq:harm_diag}. We also
omit the dependence on $k$ of $z$, $z_0$, $\omega$, $a$ for the sake of brevity, and we use the fact that $z(k)=\sqrt[4]{a(k)/b}$ (see~\eqref{eq:bogo_bos}). 
We are interested in the weakly-dissipative hydrodynamic limit $\gamma_+,\gamma_-\to 0$, 
$t\to\infty$ with fixed $\gamma_+ t, \gamma_- t$. In this limit, we can neglect some 
terms in~\eqref{eq:symbol_t_harm}, and the Fourier transformed correlator $\hat{g}_k(t)$ becomes 
\begin{widetext}
\begin{equation}
\label{eq:symbol_t_harm2}
\begin{gathered}
\hat{g}_k(t)=\begin{pmatrix}
\left(\frac{\delta_t}{z_0^2}+\frac{1-\delta_t}{g}\right)\cos^2(\omega t) + \frac{1}{z^4}\left(z_0^2\delta_t +\frac{1-\delta_t}{g}\right)\sin^2(\omega t)&\frac{2}{\omega}\left(\frac{b-a}{g}+\delta_t\left(b z_0^2+\frac{a-b}{g}-\frac{a}{z_0^2}\right)\right)\sin(2\omega t)\\
\frac{2}{\omega}\left(\frac{b-a}{g}+\delta_t\left(b z_0^2+\frac{a-b}{g}-\frac{a}{z_0^2}\right)\right)\sin(2\omega t)&\left(z_0^2\delta_t +\frac{1-\delta_t}{g}\right)\cos^2(\omega t)+z^4	\left(\frac{\delta_t}{z_0^2}+\frac{1-\delta_t}{g}\right)\sin^2(\omega t)\\
\end{pmatrix} \delta_t +\\
+\begin{pmatrix}
\frac{z^4+1}{2g z^4}(1-\delta_t) +\frac{(z^4-1) \sin(2\omega t)\delta_t}{4\omega z^4}&\frac{(z^4-1)(\cos(2\omega t)\delta_t-1)}{4z^2\omega}\\
\frac{(z^4-1)(\cos(2\omega t)\delta_t-1)}{4z^2\omega}&\frac{z^4+1}{2g}(1-\delta_t) -\frac{(z^4-1) \sin(2\omega t)\delta_t}{4\omega}\\
\end{pmatrix}.
\end{gathered}	
\end{equation}
\end{widetext}
Plugging Eq.~\eqref{eq:symbol_t_harm2} into~\eqref{eq:chbasis_harm}, we finally obtain:
\begin{widetext}
\begin{equation}
\label{eq:harm_corrt_diss}
\widetilde{\mathcal{C}}_{x,k}(t)=\mathcal{C}_{x,k}(t)\delta_t - \left(1-\delta_t\right)\times\begin{pmatrix}
\frac{(z^4+1)\delta_t}{4gz^2}+\frac{1}{2}&0&0&0\\
0&\frac{(z^4+1)\delta_t}{4gz^2}-\frac{1}{2}&0&0\\
0&0&\frac{(z^4+1)\delta_t}{4gz^2}+\frac{1}{2}&0\\
0&0&0&\frac{(z^4+1)\delta_t}{4gz^2}-\frac{1}{2}
\end{pmatrix},
\end{equation}
\end{widetext}
where $\mathcal{C}_{x,k}(t)$ is the time-dependent correlator in the absence 
of dissipation (cf.~Eq.~\eqref{eq:harm_corrt_unitary}). 
To compute the negativity carried by a shared pair, we first remove the 
oscillating phases in $\widetilde{\mathcal{C}}_{x,k}(t)$ and pass to $\widetilde{\mathcal{C}}_{x,k,t}(t)$.
Next, we change basis to obtain the position-momentum correlator $\Gamma(x,k,t)_{m,n}:=\langle\{r_m^{(k)},r_n^{(k)}\}\rangle$, 
where $r_{2j-1}^{(k)}:=x_j^{(k)}:=(\eta_j(k)+\eta^\dagger_j(k))/{\sqrt{2}}$, 
$r_{2j}^{(k)}:=p_j^{(k)}:=-i(\eta_j(k)-\eta^\dagger_j(k))/{\sqrt{2}}$, as explained in Appendix~\ref{app:chbasis} above. 
After having determined $\Gamma(x,k,t)$, we perform the partial transposition with 
respect to one of the two particles. For a system of two bosons the partial transposition of the 
density matrix is identified by  the correlation matrix $J_{\mathcal{A}_2}\Gamma J_{\mathcal{A}_2}$ (see, for instance, Ref.~\cite{Audenaert2002negativity}), 
where for a shared pair $J_{\mathcal{A}_2}$ is the matrix $J_{\mathcal{A}_2}=\mathds{1}_2\oplus\sigma_z$ (cf. Eq.~\eqref{eq:ptrans} in the main text). Thus, from~\eqref{eq:harm_corrt_diss}, we obtain 
\begin{widetext}
\begin{equation}
\label{eq:harm_corrt_gamma}
J_{\mathcal{A}_2} \Gamma(x,k,t) J_{\mathcal{A}_2}=\begin{pmatrix}
\frac{z^4+z_0^4}{2z^2z_0^2}-\frac{(z^4+1)(1-\delta_t)}{2gz^2}&0&\frac{z^4-z_0^4}{2z^2z_0^2}&0\\
0&\frac{z^4+z_0^4}{2z^2z_0^2}-\frac{(z^4+1)(1-\delta_t)}{2gz^2}&0&\frac{z^4-z_0^4}{2z^2z_0^2}\\
\frac{z^4-z_0^4}{2z^2z_0^2}&0&\frac{z^4+z_0^4}{2z^2z_0^2}-\frac{(z^4+1)(1-\delta_t)}{2gz^2}&0\\
0&\frac{z^4-z_0^4}{2z^2z_0^2}&0&\frac{z^4+z_0^4}{2z^2z_0^2}-\frac{(z^4+1)(1-\delta_t)}{2gz^2}
\end{pmatrix} \delta_t.
\end{equation}
\end{widetext}
The logarithmic negativity is obtained from the symplectic eigenvalues 
of $J_{\mathcal{A}_2}\Gamma(x,k,t)J_{\mathcal{A}_2}$, i.e., the two positive eigenvalues
\begin{align}
\label{eq:lambda1}
& \lambda_1(k,t)=\frac{z_0^2}{z^2}\delta_t + \frac{z^4+1}{2g\; z^2}(1-\delta_t), \\
\label{eq:lambda2}
& \lambda_2(k,t)=\frac{z^2}{z_0^2}\delta_t + \frac{z^4+1}{2g\; z^2}(1-\delta_t) 
\end{align}
of $i(i\sigma_y\oplus i\sigma_y) J \Gamma(x,k,t)J$, with the other two eigenvalues that are the opposite of~\eqref{eq:lambda1} and~\eqref{eq:lambda2}. Precisely, the negativity shared between the two quasiparticles of a pair is given 
as~\cite{Audenaert2002negativity} $e_2(k,t)=-\ln\left(\min\left(1,\lambda_1(k,t)\right)\right)-
\ln\left(\min\left(1,\lambda_2(k,t)\right)\right)$, which is exactly Eq.~$(16)$ 
in the main text.

\subsection{Sudden death of the negativity in the harmonic chain with gain and loss dissipation}
\label{app:sdeath}

Here we discuss the sudden death of negativity for arbitrary mass quenches in the harmonic 
chain in the presence of gain and loss dissipation. Let us first observe that the two 
eigenvalues of $i(i\sigma_y)^\oplus J_{\mathcal{A}_2} \Gamma(x,k,t) J_{\mathcal{A}_2}$ in~\eqref{eq:lambda1} and~\eqref{eq:lambda2} 
can become \textit{both} larger than $1$ at long times.  This is in constrast with the case of 
unitary dynamics, i.e., without dissipation~\cite{alba2019quantum}. This means that the negativity $e_2(k)$ of a 
pair with quasimomentum $k$ can vanish at  finite time, and the so-called sudden death of the negativity can occur. 
Interestingly, within the hydrodynamic framework, the slow entangled pairs 
with $k\to0$ play a crucial role to determine whether the sudden death can happen. 
Indeed, any pair with finite $k$ can entangle two finite intervals only for a 
finite time. At times larger than $(2\ell+d)/(2v_1(k))$, with $\ell$ the subsystems' size, and $d$ 
their distance, no pair of quasiparticles with quasimomentum $k$ can be shared between the two intervals. 
Thus, the only entangled pairs that can entangle the two subsystems at arbitrarily long times 
are the pairs with $k\to 0$. This means that if the negativity of the pairs with small 
quasimomentum $k\to0$ vanishes at finite time, the sudden death of the negativity occurs. 
Notice that for times larger than the sudden death time, the two subsystems may still be entangled, 
although their entanglement is the so-called bound entanglement~\cite{horodecki1998mixed}, 
which cannot be revealed by the negativity. 

From~\eqref{eq:lambda1} and~\eqref{eq:lambda2}, we obtain that the condition for the sudden 
death to take place if $\lambda_1>\lambda_2$ (or, equivalently, $ m_0>m_1$) is 
\begin{equation}
\label{eq:sdeath1}
|g_-| t\geq\max_{k\in [0,\epsilon]} \left|\ln\left(\frac{2gz^4-z_0^2z^4-z_0^2}{2gz^2z_0^2-z_0^2z^4-z_0^2}\right)\right|, 
\end{equation}
where $\epsilon\to0$, and we put $z_0=z_{m_0}$, $z=z_{m_1}$, with $z_m$ defined in~\eqref{eq:bogo_bos}. 
Conversely, for $\lambda_1<\lambda_2$, one has 
\begin{equation}
\label{eq:sdeath2}
|g_-| t\geq\max_{k\in [0,\epsilon]} \left|\ln\left(\frac{2gz^2-z^4-1}{2gz_0^2-z^4-1}\right)\right|.
\end{equation}
The suitable one (depending on the order between $m_0$ and $m_1$) among Eqs.~\eqref{eq:sdeath1} and~\eqref{eq:sdeath2} 
is always true for finite times, except when the logarithm diverges for $k=0$. In that case, 
the sudden death does not occur. Notice that the system can escape the sudden death only for 
$m_1w=z(0)^2=g=1$, which means $\gamma^+=0$. 
The case $g=0$, i.e. $\gamma_+=\gamma_-$, is not included in the 
conditions~\eqref{eq:sdeath1},~\eqref{eq:sdeath2}. However, it is straightforward 
to check in~\eqref{eq:lambda1} and~\eqref{eq:lambda2} that both $\lambda_1$ and $\lambda_2$  become greater than $1$ at 
finite time for any $k$, implying the presence of the sudden death in this case as well.

\end{document}